\documentclass[geosciences,review,accept,oneauthors,pdftex,10pt,a4paper]{mdpi} 
\usepackage{url}
\usepackage{ wasysym }
\usepackage{amssymb}

  \def \logg {$\log g$}
\firstpage{1} 
\pubvolume{xx}
\issuenum{1}
\articlenumber{1}
\pubyear{2019}
\copyrightyear{2019}
\externaleditor{Academic Editor: name}
\history{Received: date; Accepted: date; Published: date}

\pdfoutput=1

\usepackage{graphicx}

\Title{HEAVY METAL RULES. I.  \includegraphics[scale=0.08]{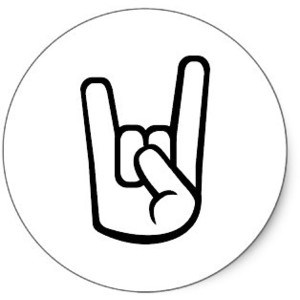}  \\ EXOPLANET INCIDENCE AND METALLICITY}

\Author{Vardan Adibekyan$^{1}$\orcidA{}}

\AuthorNames{Vardan Adibekyan}

\address{%
$^{1}$ \quad Instituto de Astrof\'isica e Ci\^encias do Espa\c{c}o, Universidade do Porto, CAUP, Rua das Estrelas, 4150-762 Porto, Portugal; vadibekyan@astro.up.pt}

\abstract{Discovery of only handful of exoplanets required to establish a correlation between giant planet occurrence and metallicity of their host stars. More than 20 years have already passed from that discovery, however, many questions are still under lively debate: What is the origin of that relation?  what is the exact functional form of the giant planet -- metallicity relation (in the metal-poor regime)?,  does such a relation exist for terrestrial planets? All these question are very important for our understanding of the formation and evolution of (exo)planets of different types around different types of stars and are subject of the present manuscript. Besides making a comprehensive literature review about the role of metallicity on the formation of exoplanets, I also revisited most of the planet -- metallicity related correlations reported in the literature using a large and homogeneous data provided by the SWEET-Cat catalog. This study lead to several new results and conclusions, two of which I believe deserve to be highlighted in the abstract: i) The hosts of sub-Jupiter mass planets ($\sim$0.6 -- 0.9~M$_{\jupiter}$) are systematically less metallic than the hosts of Jupiter-mass planets. This result might be related to the longer disk lifetime and higher amount of planet building materials available at high metallicities, which allow a formation of more massive Jupiter-like planets. ii) Contrary to the previous claims, our data and results do not support the existence of a breakpoint planetary mass at 4~M$_{\jupiter}$ above and below which planet formation channels are different. However, the results also suggest that planets of the same  (high) mass can be formed through different channels depending on the (disk) stellar mass i.e. environmental conditions.}

\keyword{exoplanets; metallicity; planetary systems; Galaxy: solar neighborhood}

\begin{document}
%%%%%%%%%%%%%%%%%%%%%%%%%%%%%%%%%%%%%%%%%%
%% Only for the journal Gels: Please place the Experimental Section after the Conclusions

%%%%%%%%%%%%%%%%%%%%%%%%%%%%%%%%%%%%%%%%%%
\setcounter{section}{0} %% Remove this when starting to work on the template.

\section{Introduction}

Understanding the origin of Earth and life on it was one of the most important and daring questions since ancient times. These questions still did not loose their actuality. Predictions about the existence of planets orbiting other stars, currently called extrasolar planets or exoplanets, have been expressed already for several centuries \citep[e.g.][]{fontenelle_entretiens_1686}. Moreover, the RV and transiting methods, the two methods that are currently the most successful ones in terms of planet detection counts and are responsible for about 95\% of all the known exoplanets, were proposed to detect these planets around other stars already 80 years ago \citep{belorizky_soleil_1938, struve_proposal_1952}. 

Starting from 1980s, thanks to the developments of high-precision spectrographs and CCD cameras, the search for exoplanets received a new impulse. Given the predictions that planets outside of our Solar System should exist numerously and the intensive search for these extrasolar planets, the detection of a planet around other stars should not have came as a surprise. However, it came: the first exoplanets\footnote{Two planets were first discovered around this neutron star in 1992 \citep{wolszczan_planetary_1992} and one additional planet was detected later in 1994 \citep{wolszczan_confirmation_1994}.} were discovered around a pulsar PSR1257 + 12 \citep{wolszczan_planetary_1992}. The existence of planets around neutron stars made a very difficult question to answer: i) did these planets form around a massive precursor star and survived the supernova explosion, ii) or they formed in a protoplanetary disk which was left over immediately after the neutron star was formed, iii) or
these planets were formed from a disk consisting of an already disrupted binary companion \citep[see][for different formation scenarios]{podsiadlowski_planet_1993}. The first confident detection of an exoplanet around a solar-type stars was made in 1995: Michel Mayor and Didier Queloz discovered a giant planet orbiting its host star with a period of only about 4 days \citep{mayor_jupiter-mass_1995}. The discovery of such a planet, which does not have an analog in our Solar System, rose more questions about formation of planets in and out of our system.  

\subsection{Diversity of exoplanet properties and their formation scenarios}

\begin{figure}[H]
\centering
\includegraphics[width=15cm]{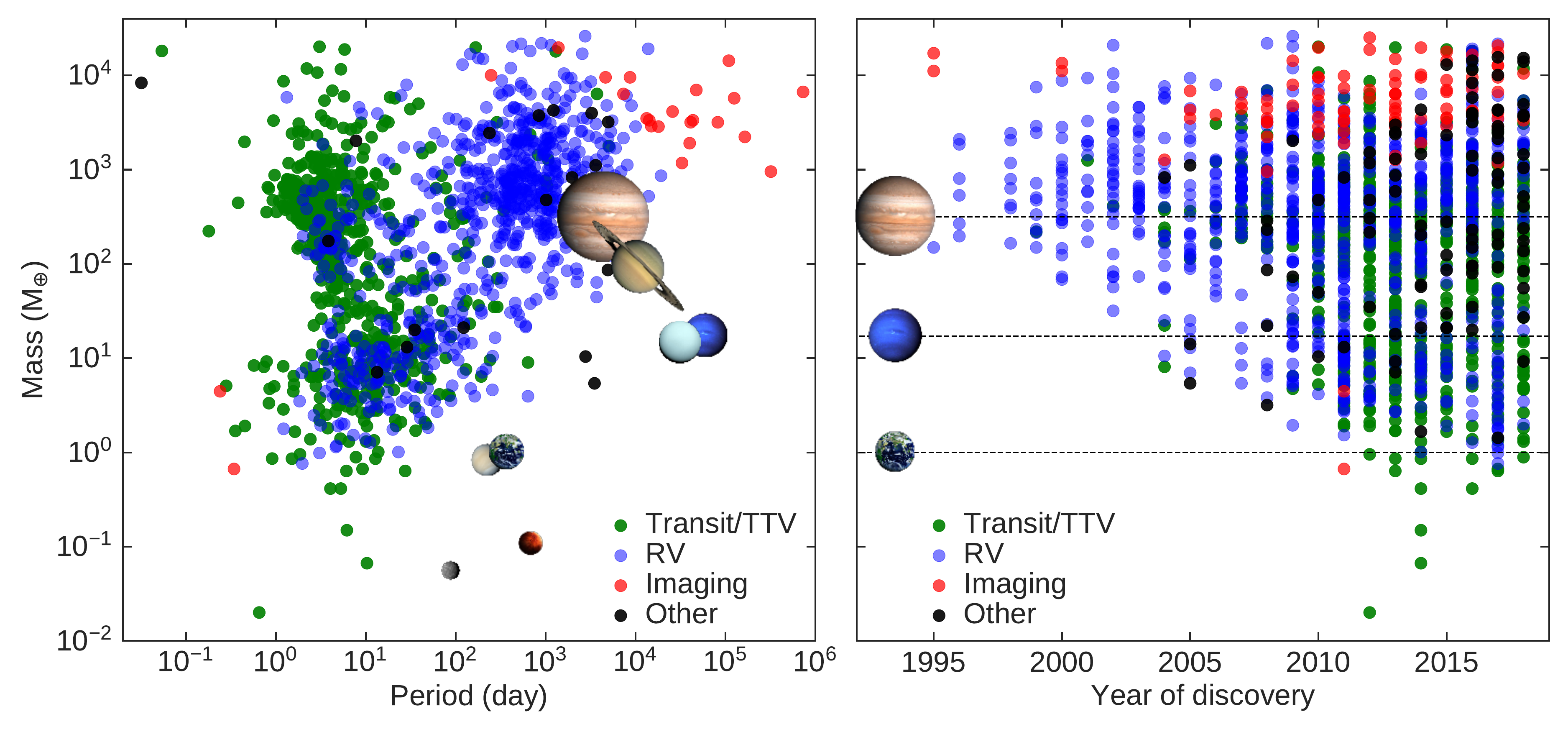}
\caption{The distribution of discovered planets (exoplanet.eu) in the period--mass diagram (left). Mass of the known planets as a function of the discovery year (right). The planets
detected by different detection techniques are shown by different symbols. The planets of our Solar System are shown as a reference.
The two low-mass and short-period planets detected by direct imaging, Kepler-70b and Kepler-70c, are found to orbit a post-red-giant star \citep{charpinet_compact_2011}.}
\label{period_mass}
\end{figure}

Since the discoveries of first exoplanets via \textit{radial velocity} \citep[51 Pegasi b ][]{mayor_jupiter-mass_1995}, \textit{transiting method} \citep[HD209458 b ][]{charbonneau_detection_2000, henry_transiting_2000}, and by \textit{direct/indirect imaging} \citep[2M1207 b/Fomalhaut b ][]{chauvin_giant_2004, kalas_optical_2008}\footnote{The ``List of exoplanet firsts'' can be found from \url{https://en.wikipedia.org/wiki/List_of_exoplanet_firsts}} almost 4000 planets have been detected\footnote{\url{http://exoplanet.eu/}} and more than 4000 candidates awaiting for validation \citep{thompson_planetary_2018}. Right panel of Fig.~\ref{period_mass} shows the masses (for planets detected by RV method only the minimum mass is available) of the so far discovered planets as a function of discovery time labeled according to detection technique. The data is extracted from Extrasolar Planets Encyclopaedia \citep{schneider_defining_2011}. Thanks to the increasing number of ground based and space based planet search surveys the number of exoplanets has increased significantly during the last years. The plot also shows that the detection of planets with masses as low as the one of Earth became possible.

Left panel of Fig.~\ref{period_mass} the distribution of exoplanets in the period -- mass diagram is plotted. The planets of our Solar System are also shown for a visual comparison. This figure suffers from selection effects and observational biases of different planet detection techniques. However, it still contains very valuable information about the processes of formation and evolution of the detected planets. In particular one can easily  see that most of the exoplanets are clustered in three groups: i) hot-Jupiters with M $_{p}$ $\sim$1-2~M$_{\jupiter}$, ii) gas and ice giants with~M$_{p}$ $\sim$1-2~M$_{\jupiter}$ and P $\sim$1000 days, and iii) hot super-Earths with~M$_{p}$ $\sim$10~M$_{\oplus}$ and P $<$ 100 days. The observed lack of giant planets with periods between 10-100 days, so called ``period-valley'' \citep[][]{udry_statistical_2003, santerne_sophie_2016} is probably related to the disk migration \citep[see][for a discussion]{dawson_origins_2018} and the relative paucity of intermediate mass planets with masses between 10 to 100 M$_{\oplus}$ (so called ``planetary desert'' \citep[e.g.][]{ida_toward_2004})\footnote{The recent surveys of microlensing planets suggest about 10 times more intermediate-mass (20 -- 80 M$_\oplus$ for a median host star mass of 0.6 M$\odot$) planets than the CA models of \citet{ida_toward_2004} and \citet{mordasini_extrasolar_2009} predict  \citep{suzuki_microlensing_2018}.} is explained as a consequence of the rapid increase in mass through runaway accretion when the planet core and envelope mass reach to about 30M$_{\oplus}$ \citep[e.g.][]{ida_toward_2004, mordasini_harps_2011}. Note, that the recent CA population synthesis of \citet[][]{mordasini_planetary_2018} suggest not as strong ``planetary desert'' as was previously proposed \citep[e.g.][]{ida_toward_2004}. The reason that the planets of our SS occupy the empty regions of the diagram is partly due to detection biases. At the same time, in our SS there are no short period (shorter than about 100 days) super-Earths\footnote{I note that using the term 'super-Earth' can be misleading or confusing for a reader (especially for a reader outside of the field or for public audience) \citep{moore_how_2017}. That is not our intention. By super-Earth I mean (as almost everywhere in the literature) planets more massive or larger than the Earth. I do not imply that more information about the properties of these planets are known that makes them similar to the Earth.}, which are very common around other stars. This property makes our SS somehow special \citep{martin_solar_2015}\footnote{It was also proposed that Jupiter's periastron is atypical making the  SS an outlier \citep[e.g.][]{beer_how_2004}. However, the Jupiter's  'atypical' periastron is probably due to the observational biases \citep{martin_solar_2015}.}.  For more details about the distribution of planets on the period -- mass diagram and on the architecture of exoplanetary systems I refer the reader to the recent excellent review by \citet{winn_occurrence_2015}.

\begin{figure}[H]
\centering
\includegraphics[width=14cm]{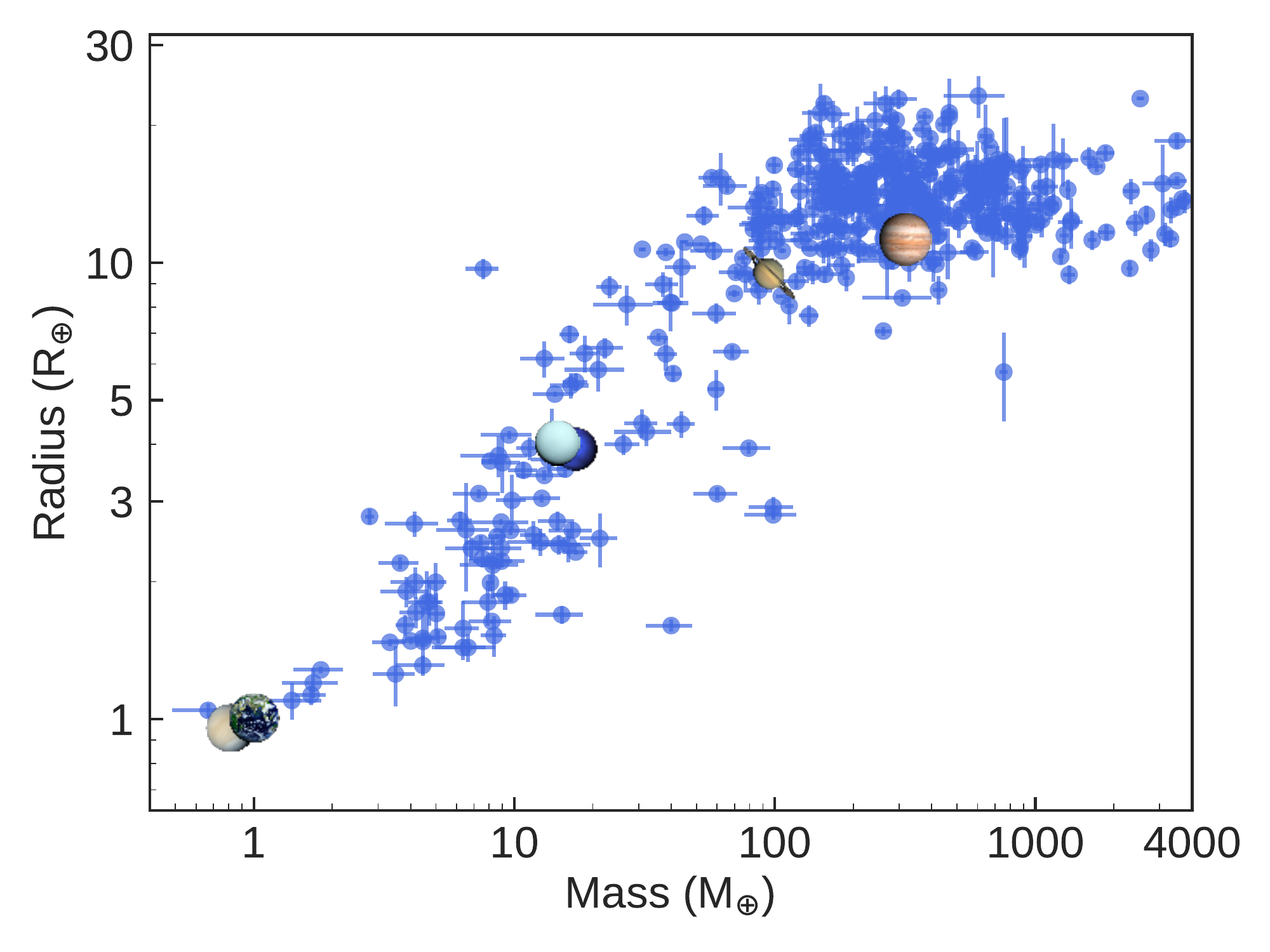}
\vspace{-0.3cm}
\caption{The mass-radius relation  for the discovered exoplanets (exoplanet.eu) with mass and radius determination better than 30\%.
The planets of our Solar System are shown for comparison.}
\label{radius_mass}
\end{figure}

Fig.~\ref{radius_mass} shows the mass--radius diagram of the extrasolar planets with precise measurements of mass and radius better than 30\%. Similar to Fig.~\ref{period_mass} I see large diversity of the physical characteristics of these exoplanets. In particular it is apparent the large number of Jupiter mass planets with radii much larger than that of Jupiter. Most of these bloated planets have very short orbits receiving incident flux of 2-3 magnitudes higher than the Earth receives \citep{demory_lack_2011, miller_heavy-element_2011}.  Several theoretical explanations have been proposed \citep{bodenheimer_radii_2003, guillot_evolution_2002, chabrier_heat_2007} that can efficiently explain the radii of individual planets, but they have difficulties to explain the properties of  the entire population of these puffed-up planets. The recent work by \citet{sestovic_investigating_2018} suggests that the inflation extent  depends on the mass of  the planets. In the figure it is also clear that there is a transition mass at which the mass-radius relation changes its functional form. Statistical studies of the mass--radius relation can provide valuable information about the bulk composition and structure of these planets, and even information about their formation and evolution  \citep{seager_massradius_2007, fortney_planetary_2007, hatzes_definition_2015}. 

\begin{figure}[H]
\centering
\includegraphics[width=11cm]{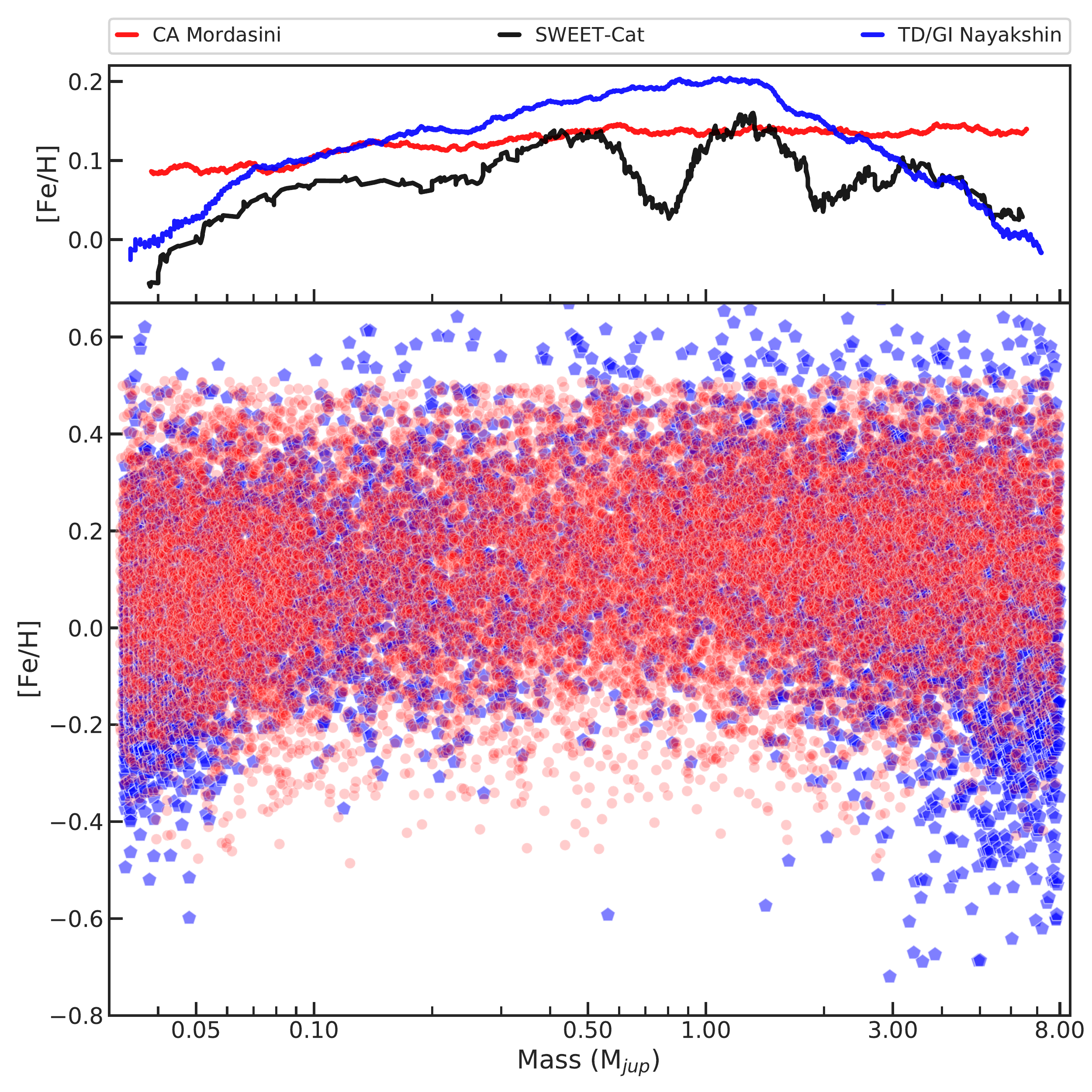}
\vspace{-0.3cm}
\caption{(Bottom panel): Dependence of the mass of synthetic planets predicted by the CA \citep{mordasini_extrasolar_2012} and TD/GI  \citep{nayakshin_tidal_2016} planet formation models against the metallicity of their host stars. The running mean of [Fe/H] as a function of mass of CA planets (in red), TD/GI planets (in blue), and observed planets (in black) is shown in the top panel. The running means were calculated using windows of 50, 500, and 1000 for observed, TD/GI, and CA planets, respectively. These numbers reflect the sizes of each sample.}
\label{nayakshin_mordasini}
\end{figure}

The radius-mass (Fig.~\ref{period_mass}) and period-mass (Fig.~\ref{radius_mass}) diagrams (among other parameter spaces) contain enormous information about the physical processes acting during the formation and evolution of (exo)planets. Thus the reproduction of these diagrams is the ultimate goal of any planet formation theory. The models developed to explain the formation of SS planets did not predict such an enormous diversity of exoplanets  that we observe  in Figs.~\ref{period_mass} and ~\ref{radius_mass} \citep[e.g.][]{boss_proximity_1995, lin_orbital_1996}. The two most commonly used exoplanet formation scenarios are the core accretion \citep[CA; e.g.][]{safronov_evoliutsiia_1969, pollack_formation_1996, ida_toward_2004, alibert_models_2005, mordasini_extrasolar_2009, hasegawa_origin_2011} and gravitational instability \citep[GI; e.g.][]{kuiper_origin_1951, boss_evolution_1998, boss_stellar_2002, boley_clumps_2010, vorobyov_gravitational_2018}. In the CA scenario the low-mass planets and cores of massive planets form from the coagulation of small grains in the protoplanetary disc. When these cores reach to a critical mass of about 5-10M$_{\oplus}$ \citep{ikoma_formation_2000, hasegawa_planet_2014, mordasini_grain_2014} they undergo runaway accretion of gas. Giant planets then will be formed if the critical core mass is reached before the dissipation of the protoplanetary disk \citep[e.g.][]{tychoniec_vla_2018}, which has a typical lifetime of several Myr \citep{haisch_disk_2001, mamajek_initial_2009}. Since the phase of runaway accretion is very fast the planet formation time is essentially defined by the time of core-formation \citep[e.g.][]{alibert_models_2005}. In the GI scenario, the massive and cold disks fragment into a few Jupiter mass clump which then contracts to form giant planets \citep[e.g.][]{boss_giant_1997}. These massive planets are formed quickly, much before the gas in the disk depletes \citep[see][and references therein]{durisen_gravitational_2007}. If the GI is followed by tidal stripping during the migration of the clumps, then even a rocky planet (including objects in the SS \citep{nayakshin_tidal_2011, nayakshin_differentiation_2014}), can be formed \citep[see][for a recent 
review on ``Tidal Downsizing'']{nayakshin_dawes_2017}.

Both the classical CA and GI theories have experienced substantial development and modifications during the last decades \citep[e.g.][]{alessi_formation_2018, johansen_forming_2017, boss_effect_2017, nayakshin_tidal_2015}. They include important processes such as pebble accretion \citep[e.g.][]{johansen_rapid_2007, ida_radial_2016, alibert_formation_2018}  and/or migration in the disk \citep[e.g.][]{alibert_migration_2004}. Both theories have their strengths and weaknesses \citep[][]{helled_giant_2014} and might operate efficiently under different physical conditions and different parameter space \citep{matsuo_planetary_2007}.  The planetary population synthesis calculations \citep{ida_toward_2004, mordasini_global_2015, forgan_towards_2013, ndugu_planet_2018} based both on recent CA \citep[e.g.][]{mordasini_extrasolar_2009,hasegawa_planetary_2013, ndugu_planet_2018}  and TD \citep[e.g.][]{nayakshin_dawes_2017, forgan_towards_2018} are already able to reproduce some of the main structures of Figs.~\ref{period_mass} and \ref{radius_mass}.

In the bottom panel of Fig.~\ref{nayakshin_mordasini} I plot the mass of synthetic planets formed through CA of \citet{mordasini_extrasolar_2012}\footnote{The data is downloaded from \url{http://www.mpia.de/~mordasini/Site7.html}.} and through TD/GI of \citet{nayakshin_tidal_2016}\footnote{The data is kindly provided by Sergei Nayakshin.} against the metallicity of their host stars. The plot shows that the two conceptually different models cover very similar areas in this diagram. On the top panel of the figure I plot running mean of the host metallicity as a function of planetary mass for CA synthetic planets, TD/GI synthetic planet, and confirmed planets with hosts metallicities derived in a homogeneous way in SWEET-Cat (see Sect.~\ref{sweetcat}). This panel clearly shows that when one looks at the populations of planets as a whole, the two models show significant deviation from each other and from the observed planet populations in some metallicity regions. Although in the two considered models the absolute values of the mean [Fe/H] may change with population synthesis parameters, the trends should still remain since they are based on solid physical principles. Thus testing the predictions of the planet formation models with observations is the way to identify the weaknesses in the models and to provide insightful information for their further development.

\section{Motivation and the outline of this review}

As discussed in the previous section, the modern planet formation theories need to consider many different physical phenomena (e.g. migration of planets, evaporation, pebble accretion) in order to reproduce the main properties of the detected exoplanets. Besides these physical processes that directly affect the characteristics of the planets and their 
orbital architecture, one should also consider the environmental conditions where these planets are formed \citep[e.g.][]{adibekyan_formation_2017}. 

Linking the properties of exoplanets with the properties of protoplanetary disks where they have been formed will provide important insights about planet formation in different environments. Studying the dependencies of exoplanets properties on environmental conditions are usually performed indirectly by looking at the physical properties of the host stars that are linked to the characteristics of the proto-stellar/planetary disk. 
Since the planets and their host stars are formed from the same molecular clouds, some global environmental properties such us chemical composition of some important mineralogical abundance ratios \citep[e.g. Mg/Si, Fe/Si][]{bond_compositional_2010, thiabaud_stellar_2014, dorn_can_2015}\footnote{Note that the C/O ratio, that controls the amount of the carbides and silicates in planet building blocks, varies in the protoplanetary disk and might be different from the value of the parent star  \citep[][]{madhusudhan_toward_2014, thiabaud_gas_2015}.}, can be determined from the spectroscopic studies of their host stars. Moreover, precise characterization of the planet host stars is crucial for the characterization of the planets themselves \citep{adibekyan_characterization_2018}.

In this manuscript I will review the main observational correlations between the properties of exoplanets and the metallicity of their host stars. I  will discuss the most-studied correlation between occurrence rate of different types of planets with metallicity (iron content) of their host stars and its link to different planet formation processes and models. Sine some of the correlations reported in the literature are based on heterogeneous data (for example compilation of properties of hosts stars from different literature sources), whenever possible, I will revisit them using the largest homogeneous catalog of exoplanet host stars called SWEET-Cat \citep[see Sect.~\ref{sweetcat},][]{santos_sweet-cat:_2013}.

\subsection{SWEET-Cat} \label{sweetcat}

There are four exoplanet catalogs that exoplanetologists commonly use \citep{bashi_quantitative_2018}. Despite small biases and discrepancies between these catalogs, on average they provide a good compilation of properties of exoplanets and their host stars \citep{bashi_quantitative_2018}. All these catalogs, however, suffer from heterogeneous literature compilations of stellar parameters. This heterogeneous compilation makes significant discrepancies when comparing with parameters derived in a homogeneous way \citep[e.g.][]{sousa_homogeneous_2015}. Moreover, homogeneous and uniform spectral analysis is essential to minimize the uncertainties in stellar atmospheric properties \citep[e.g.][]{torres_improved_2012}.

\begin{figure}[H]
\centering
\includegraphics[width=15cm]{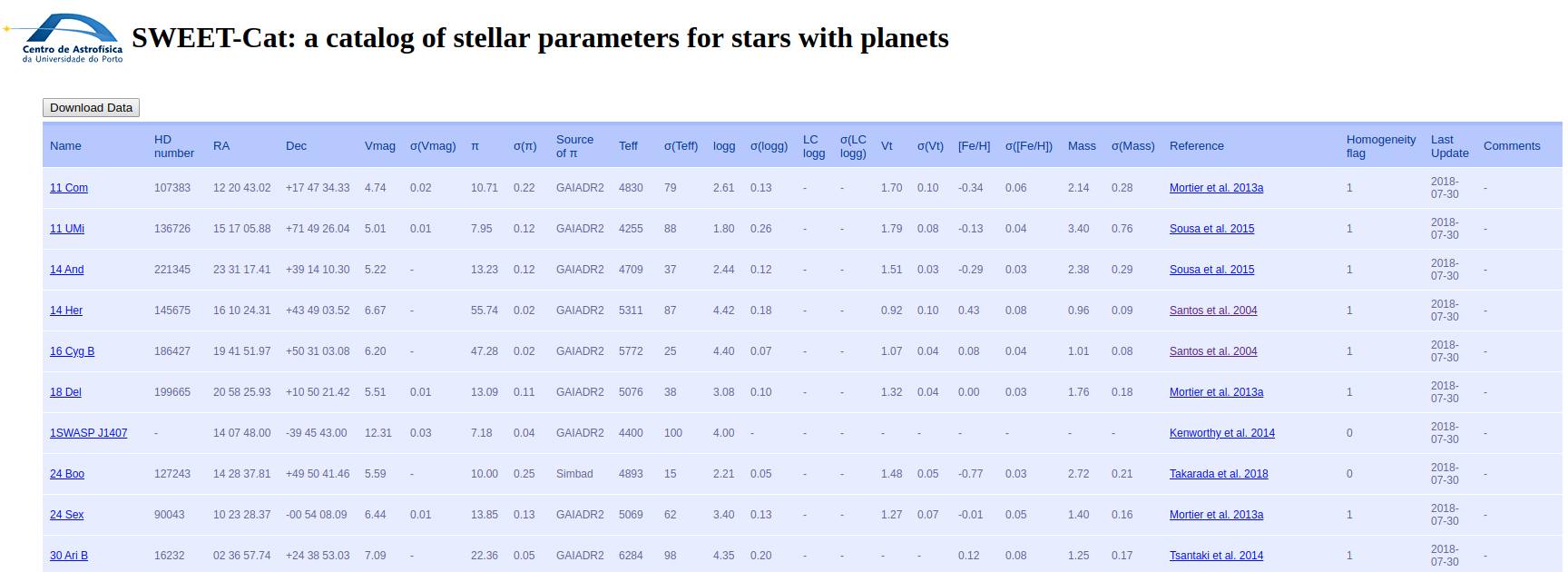}
\vspace{-0.2cm}
\caption{Screenshot showing the structure of the SWEET-Cat.}
\label{fig-sweetcat}
\end{figure}

\citet{santos_sweet-cat:_2013}, presented a catalog of stellar parameters for stars with planets (SWEET-Cat\footnote{https://www.astro.up.pt/resources/sweet-cat/}) listed in the Extrasolar Planets Encyclopaedia \citep[exoplanet.eu][]{schneider_defining_2011}. This catalog provides a compilation of atmospheric stellar parameters (see Fig.~\ref{fig-sweetcat}) from literature and, whenever possible, derived using the same uniform methodology \citep[see e.g.][]{santos_spectroscopic_2004,sousa_spectroscopic_2008, sousa_ares_2014}. The catalog is regularly updated \citep[][]{andreasen_sweet-cat_2017, sousa_homogeneous_2015} and after the last major update \citep{sousa_sweet-cat_2018} it contains uniformly derived stellar parameters of $\sim$80\% of bright stars ($V<$12 mag) hosting RV detected planets. Besides the main stellar parameters, SWEET-Cat will soon provide abundance of different chemical abundances of the host stars again derived in a homogeneous and uniform way [Sousa et al. in prep.]. 

It is worth to note that different groups also intensively work on homogeneous derivation of stellar parameters of Kepler stars hosting planets and planet candidates e.g. The Kepler Follow-up Observation Program \citep[][]{furlan_kepler_2018} and The California-Kepler Survey \citep[][]{petigura_california-kepler_2017}. Moreover, during the past few years dedicated communities\footnote{SAG-14 (https://exoplanets.nasa.gov/exep/exopag/sag/).} and web interfaces\footnote{ ExoFOP (https://exofop.ipac.caltech.edu/).} have been created to optimize the resources in exoplanet follow-up studies and characterization of their host stars.

\subsection{Nomenclature: the term 'metallicity' in astronomy} \label{nomenclature}

The term 'metallicity' is very widely used in the literature. Although it is always used to quantify the amount of 'metals', its nomenclature and scales are not the same in different fields of astronomy. In this subsection I will very briefly present the definitions of metallicity, its scales, and the quantities that are derived in practice as a proxy of metallicity in different fields of stellar astronomy.

\paragraph{Mass fraction}

In astronomy, all the chemical elements heavier than hydrogen and helium are called metals\footnote{This definition should not be confused with the definition of metals in the period table: \url{https://en.wikipedia.org/wiki/Properties_of_metals,_metalloids_and_nonmetals}.}. The composition of the stars can be characterized by the mass fraction of hydrogen (denoted as $X$), helium (denoted as $Y$), and metals (denoted as $Z$). The sum of these three abundances is normalized to unity i.e. $X$ + $Y$ + $Z$ = 1. This definition is convenient when dealing with the models of stars. However, direct derivations of these quantities from observations is impossible. Note, that in the literature $Z$ is sometimes called ``metal mass fraction'' \citep[e.g.][]{nsamba__2018} and sometimes ``metallicity'' \citep[e.g.][]{bressan_parsec:_2012}.

\paragraph{The solar ($\log \varepsilon(H) \equiv 12$) scale}

Atomic abundances of different metals (elements with atomic number greater than 2) in stars vary over several magnitudes, and thus, it is convenient to express them in a logarithmic scale. In stellar astrophysics, the atomic abundances of elements are usually presented relative to H\footnote{Note that for example the meteoritic abundances are usually pegged to Si abundance which is fixed to $N_{Si} \equiv 10^{6}$.}. Traditionally the logarithmic abundance (number of atoms) of H is set to 12 i.e. $\log \varepsilon(H) \equiv 12$. Atomic abundance of any other element El, then will be

$$ \log \varepsilon(El) = \log \bigg(\frac{N_{El}}{N_{H}}\bigg) + 12$$

where $N_{El}$ and $N_{H}$ are the number of atoms of element X and H ($\log (N_{H})$ = 12). The unit used for the abundances is the ``dex'' which is  contraction of ``decimal exponent''. 

Sometimes it is more convenient to derive the atomic abundances of stars relative to  Sun using the $\log \varepsilon(H) \equiv 12$ scale. In this case square braces are used to express the abundance of an element El

$$ [El/H]  = \log \bigg(\frac{N_{El}}{N_{H}}\bigg)_{star} -  \log \bigg(\frac{N_{El}}{N_{H}}\bigg)_{sun} $$

Obviously for the Sun [El/H] $\equiv$ 0 for any element. Following this definition, abundance ratio of any two elements El1 and El2 can be expresses as 

$$ [El_{1}/El_{2}] = \log \bigg(\frac{N_{El_{1}}}{N_{El_{2}}}\bigg)_{star} -  \log \bigg(\frac{N_{El_{1}}}{N_{El_{2}}}\bigg)_{sun}$$

The conversion from solar scaled to absolute abundances requires the knowledge of the reference solar abundances  \citep[e.g.][]{lodders_solar_2003, asplund_chemical_2009, lodders_abundances_2009, caffau_solar_2011}.

Metallicity of a star [m/H] is related to the metal mass fraction Z in the following form

$$ [m/H] = \log \bigg(\frac{Z}{X}\bigg)_{star} -  \log \bigg(\frac{Z}{X}\bigg)_{sun}$$

Unfortunately, it is very difficult to derive atomic abundances of all the  metals for a given star. Thus, in stellar astrophysics, the iron content ([Fe/H]) is commonly used as a tracer of overall stellar metallicity. 
Although iron is not the most abundant metal in the Universe, the visible spectra of solar-type stars contain many strong iron lines which are easy the measure\footnote{In nebular astrophysics oxygen abundance is usually 
used as the standard scale. This is simply because oxygen emission lines are among the strongest ones in nebular spectra.}. The usage of iron abundance as a proxy of overall meallicity assumes that the abundances of all the metals change proportionally to iron content.  Thus under the assumption that the distribution of heavy metals in a given star is the same as in the Sun $\{X_{i}/Z\}$ = $\{X_{i}/Z\}_\odot$ (solar-scaled distribution) one can write 

$$ [m/H] = [Fe/H] = \log \bigg(\frac{Z}{X}\bigg)_{star} -  \log \bigg(\frac{Z}{X}\bigg)_{sun}$$

Taking into account the solar abundance \citet{bertelli_theoretical_1994} suggests the following approximate relation between [Fe/H] and Z \citep[see also][]{bonfanti_age_2016}

$$ \log(Z) = 0.977[Fe/H]-1.699 \quad \textrm{or} \quad  [Fe/H] = 1.024\log(Z) + 1.739$$

The assumption about the universality of solar-scaled heavy element distribution is not always valid, especially for metal-poor stars that are enhanced in $\alpha$ elements \citep[e.g.][]{fuhrmann_nearby_1998, reddy_elemental_2006, adibekyan_chemical_2012, adibekyan_kinematics_2013, recio-blanco_gaia-eso_2014}. To account for the $\alpha$-enrichment the following correction is proposed \citep{yi_toward_2001}

$$ [m/H]  = [Fe/H] + \log(0.694f_{\alpha} + 0.306) \quad \textrm{where} \quad  f_{\alpha} = 10^{[\alpha/Fe]} $$

For a recent review on the derivation of stellar metallicities and other spectroscopic parameters of exoplanet hosting stars I refer the reader to \citet{adibekyan_characterization_2018}.

\section{Metallicity of planet host stars} \label{pl_occur_metal}

The first observed correlation that linked the presence of planets and a property of their host stars was the giant planet -- stellar metallicity correlation \citep[e.g.][]{gonzalez_stellar_1997,santos_metal-rich_2001}. The discovery of this  correlation played a  crucial role for the advancement of planet formation theories \citep[e.g.][]{ida_toward_2004, mordasini_extrasolar_2012, nayakshin_dawes_2017}.  Until now, studies of the possible correlations between different types of planets (e.g. low-mass or massive planets at short or long period orbits) and metallicities of stars of different spectral types and evolutionary stages intensively continue, and provide new constraints for the models of planet formation and evolution. Planet--metallicity correlation is the subject of the discussion of this section. 

It is very important to note, that in this manuscript the iron content is used as a proxy of overall metallicity\footnote{It is important to remember that the stellar chemical abundances and metallicities are usually derived from photospheric absorption lines thus they represent the composition of the photosphere.}. This approximation is very commonly used in the literature and is well justified for solar metallicity stars. However, as already mentioned, at low metallicities the iron abundance does not necessarily equal to the overall metal content in stellar atmospheres. There is another limitation that is typically ignored when studying the metallicity distribution of planet hosting stars. There are some astrophysical processes, such as atomic diffusion due to concentration and thermal gradients \citep[e.g.][]{mowlavi_stellar_2012}, that can affect the stellar metallicity. Moreover, engulfment of planetisimals under specific conditions \citep[e.g. depending on the number of accretion episodes][]{theado_metal-rich_2012} and/or non-accretion of metal-rich material as a consequence of planet formation \citep[again, depending on the time of planet formation and the composition of the formed planets][]{kunitomo_revisiting_2018} may also change the original metallicity of the planet hosting stars. Thus, the present-day stellar metallicity can be slightly different from the metallicity of the stars at the time of their formation \citep[e.g.][]{asplund_chemical_2009}.. Finally, it is important to note, that the key parameter in the core-accretion models of planet formation is the solid surface density \citep[e.g.][]{ida_toward_2004, mordasini_extrasolar_2009}. This parameter is proportional to the dust-to-gas ratio in the disk and is very difficult to constrain by observations. In turn, average dust-to-gas ratio in the protoplanetary disks is proportional to the metal abundance of the  ISM \citep[e.g.][]{ruffle_galactic_2007} and is usually assumed to be proportional to the atmospheric metallicity of the host stars \citep[e.g.][]{murray_stellar_2001,mordasini_extrasolar_2009, ercolano_metallicity_2010}. The validity of this assumption was very recently revisited by \citet[][]{liu_migration_2016}. The authors concluded that although the metallicity of the disk can change significantly as the disks evolve, the dust-to-gas ratio (within an order of magnitude) is comparable with the heavy element content in the atmosphere of the host stars. However, for particular cases and in particular locations the dust-to-gas ratio can be significantly different from the host star metallicity \citep[e.g.][]{dawson_metallicity_2015} because of particles being decoupled from gas and radially drifting inwards \citep[e.g.][]{andrews_tw_2012, birnstiel_outer_2014}.

\subsection{Giant planets and metallicity} \label{giant_planet_metal}

Detection of only four giant planets required from \citet{gonzalez_stellar_1997} to notice that these planets tend to appear around metal-rich stars. This hint was very soon confirmed using relatively larger samples \citep[e.g.][]{laughlin_mining_2000, santos_metal-rich_2001, gonzalez_parent_2001} and become a well established correlation \citep[e.g.][]{santos_spectroscopic_2004, fischer_planet-metallicity_2005, johnson_giant_2010, sousa_spectroscopic_2011, mortier_functional_2013} known as giant planet -- metallicity correlation.

At the end of 1990s and beginning of 2000s, when only handful of hot-Jupiters have been detected, two hypotheses have been proposed to explain the metal excess of massive exoplanet host stars: self enrichment (aka pollution) mechanism and primordial origin \citep[e.g.][]{gonzalez_stellar_1997}. In the self enrichment scenario the high metallicity of planet hosts was explained by the pollution of outer convective envelope of the hosts due to accretion of gas-depleted material. This pollution could be a result of inward migration of massive planets that sweep metal-rich material with them. In the alternative, primordial scenario,  the over-abundance of metals in massive planet hosts was considered to represent the high metallicity of the primordial cloud where the stars have been formed and was assumed that giant planets form more readily in high metallicity environments.  

The two aforementioned scenarios provide different observationally testable signatures for the host stars. In particular, in case of pollution by rocky material, the increase of atmospheric metallicity of the host stars should depend on the mass of convective zone, thus on the mass and temperature of the stars \citep[][]{gonzalez_stellar_1997, laughlin_possible_1997, pinsonneault_mass_2001, murray_stellar_2001}. Going one step further, \citet{gonzalez_stellar_1997} proposed that if the accreting material is fractionated then a trend between chemical abundances and condensation temperature of the elements is expected. \citet{gonzalez_stellar_1997} also noted that the high-metallicity accretion signature should be lost for evolved stars which have larger convective zones where the accreted heavy metals will be mixed and diluted. When the number of detected planets increased, several systematic abundance studies on stars with and without planets have been performed focused on the aforementioned predictions. The results of these studies   favored the primordial cloud as the most likely origin for the metal-rich nature of giant planet host stars \citep[e.g.][]{santos_spectroscopic_2004, fischer_planet-metallicity_2005, valenti_stellar_2008, johnson_giant_2010}. This hypothesis is supported by models of planet formation and evolution based on CA \citep[e.g.][]{ida_toward_2004, mordasini_extrasolar_2012} and tidal downsizing \citep{nayakshin_tidal_2015}.

\begin{figure}[H]
\centering
\includegraphics[width=13cm]{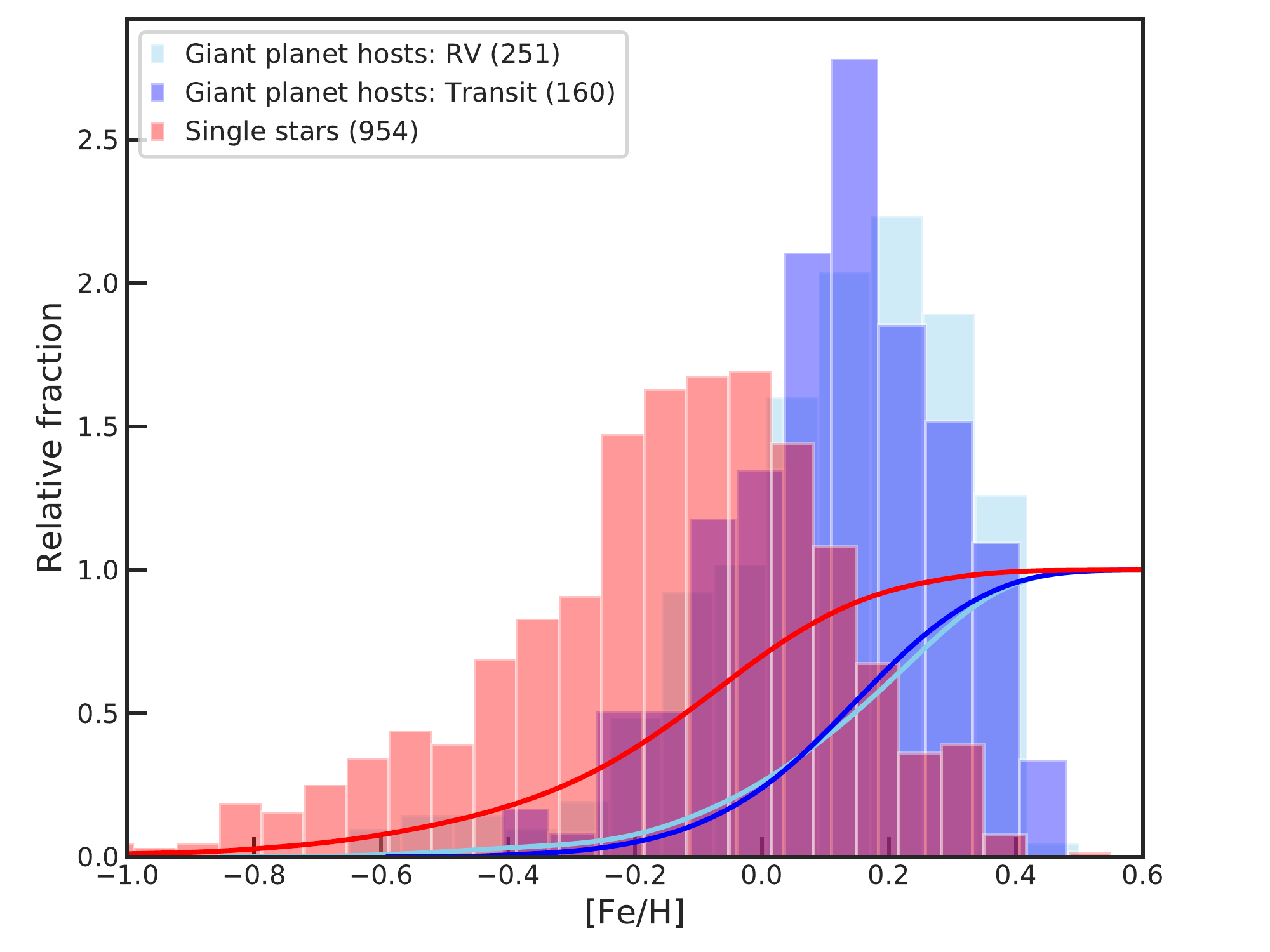}
\vspace{-0.2cm}
\caption{Metallicity distribution of stars hosting giant planets detected via RV method (skyblue) and Transit method (blue), and stars without any detected planets (red). The KDE fit of the cumulative distribution of [Fe/H] for the three sample of stars is also shown with the curves of corresponding colors.}
\label{fig-giant_planet_metal}
\end{figure}

In Fig.~\ref{fig-giant_planet_metal} I plot the metallicity distribution of stars hosting high mass planets (hereafter, HMPH) and stars known to host no detected planet (hereafter, SnoP). Only FGK dwarf hosts (\logg \ $>$ 4.0 dex, and 0.6~M$_{\odot} <$ M $<$ 1.5~M$_{\odot}$)   stars with  with stellar parameters derived in a homogeneous way in SWEET-Cat are considered. The comparison sample of stars without detected planets consists of 954 FGK dwarf stars observed within the HARPS GTO program and are taken from \citet{adibekyan_chemical_2012}. In this figure only planets with masses between 50~M$_{\oplus}$ and 13~M$_{\jupiter}$\footnote{13M$_{\jupiter}$ approximately corresponds to the  deuterium-burning mass limit for BDs at solar metallicity \citep[e.g.][]{spiegel_deuterium-burning_2011} and is usually used as a border line between giant planets and BDs. The exact deuterium-burning mass limit depends on several parameters \citep[see e.g.][for more details]{chabrier_giant_2014, caballero_review_2018}.} are  considered. In case of multiplanetary systems the mass of the most massive planet is considered. I  note that there is no clear definition for what should be the low mass limit of giant planets, thus the choice for lower limit of mass M = 50~M$_{\oplus}$ is somewhat semi-arbitrary. Different authors chose different definitions\footnote{I  refer the readers to \citet{chabrier_giant_2014} and \citet{schneider_defining_2011} for an interesting discussion about the classifications of planets and BDs based on physical observable properties and formation mechanisms.} for giant planets and the choice of the low-mass limit varies (typically between 30 to 90~M$_{\oplus}$) depending on the aim of the study \citep[e.g.][]{cumming_keck_2008,russell_geophysical_2013, hatzes_definition_2015, brucalassi_search_2016, pinotti_zero_2017, bashi_two_2017, sousa_sweet-cat_2018}. In any case I choose this limit since it is close to the mass 'gap' observed in Fig.~\ref{period_mass} and \ref{radius_mass}.  At the same time the planet core-accretion based population synthesis models by \citet{mordasini_extrasolar_2009} predicted such a minimum in the mass-distribution at about this  mass. It is important to note that presented results are not sensitive to the choice of these limits. Here I should also comment that the classifications of different types of planets based on their radii is not straightforward as well, and varies from work to work \citep[e.g.][]{buchhave_three_2014, fulton_california-kepler_2017, narang_properties_2018, petigura_california-kepler_2018, berger_revised_2018}.

Although some theoretical works based on CA models propose that the critical metallicity for giant planet formation is in the range of $-$1.8 to $-$1.5 dex\footnote{Note that GI based model of \citet{johnson_constraints_2013}  suggests a critical metallicity of about $-$4 dex for planet formation to happen.} \citep[][]{hasegawa_planet_2014, johnson_first_2012}, the lowest metallicity of the giant planet host in Fig.~\ref{fig-giant_planet_metal} is -0.65 dex. In fact, the lowest metallicity of a confirmed sub-stellar companion host currently listed in exoplanet.eu is $-$1.00$\pm$0.07 dex derived for BD+202457\footnote{BD+202457 is known to host two companions with masses of 12.47~M$_{\jupiter}$ and 21.42M$_{\jupiter}$ \citep{niedzielski_substellar-mass_2009}.} \citep{niedzielski_substellar-mass_2009}.  However, the more recent spectroscopic analysis of this star suggests a higher  metallicity of [Fe/H] = $-$0.79 \citep{maldonado_metallicity_2013}. Besides BD+202457, the lowest (spectroscopically derived)  metallicity star hosting a giant planet in exoplanet.eu is the 24 Bo\"{o}tis with a metallicity of $-$0.77$\pm$0.03 dex \citep[][]{takarada_planets_2018}. Observationally determining the metallicity limit below which giant planets do not form can provide a very important insights for the planet formation theories. Indeed, some programs are focused on the metal-poor  regime trying to tackle this issue \citep[e.g.][]{sozzetti_keck_2009, santos_harps_2011, mortier_frequency_2012}.

From Fig.~\ref{fig-giant_planet_metal} it is clear that the SnoP and HMPH samples have different metallicity distributions, the latter ones being more metallic. In particular the mean metallicity of SnoP is $-$0.159$\pm$0.009 dex, while the average metallicity of the stars hosting high mass transiting and RV planets are 0.117$\pm$0.013 dex and 0.112$\pm$0.013 dex, respectively. Here the errors represent the standard error of the mean i.e.  standard deviation divided by the square root of the sample size. To evaluate the significance of this difference more quantitatively I applied a two--sample KS test to the samples. The results presented in Table~\ref{table_KS_metal_HMPH} show that stars hosting giant planets have significantly different metallicity distribution when compared to the stars without planets. Interestingly, the two samples of HMPH of transiting and RV planets have metallicity distributions that come from the same parent distribution. This is somehow surprising  since the distributions of the orbital periods of the transiting and RV planets are statistically different. The transiting planets orbit their stars on average in 11 days (std = 66 days), while the RV detected planets have an average orbital period of 1202 days (std = 1775 days), and the KS tests suggests a p-value of 3.5$\times10^{-61}$ for the two orbital period distributions being similar.

As already mentioned the giant planet -- metallicity relation  provides a general but very important information for planet formation models. The knowledge of the exact functional form of this relation can provide additional important constraints for the existing planet formation theories. However, a large, unbiased, and volume-limited sample of stars surveyed for planets is required to study the details of this relation. Usually, the dependence of planet fraction ($f$) on stellar metallicity is described by the following functional form: $f$ = $\alpha 10^{\beta [Fe/H]}$, where $\alpha$ and $\beta$ are the coefficients to be derived. To include the dependence of planet occurrence rate on stellar metallicity usually the $f([Fe/H])$ is fitted with the functional form of $f$ = $\alpha 10^{\beta [Fe/H]}$.

\begin{figure}[H]
\centering
\includegraphics[width=14cm]{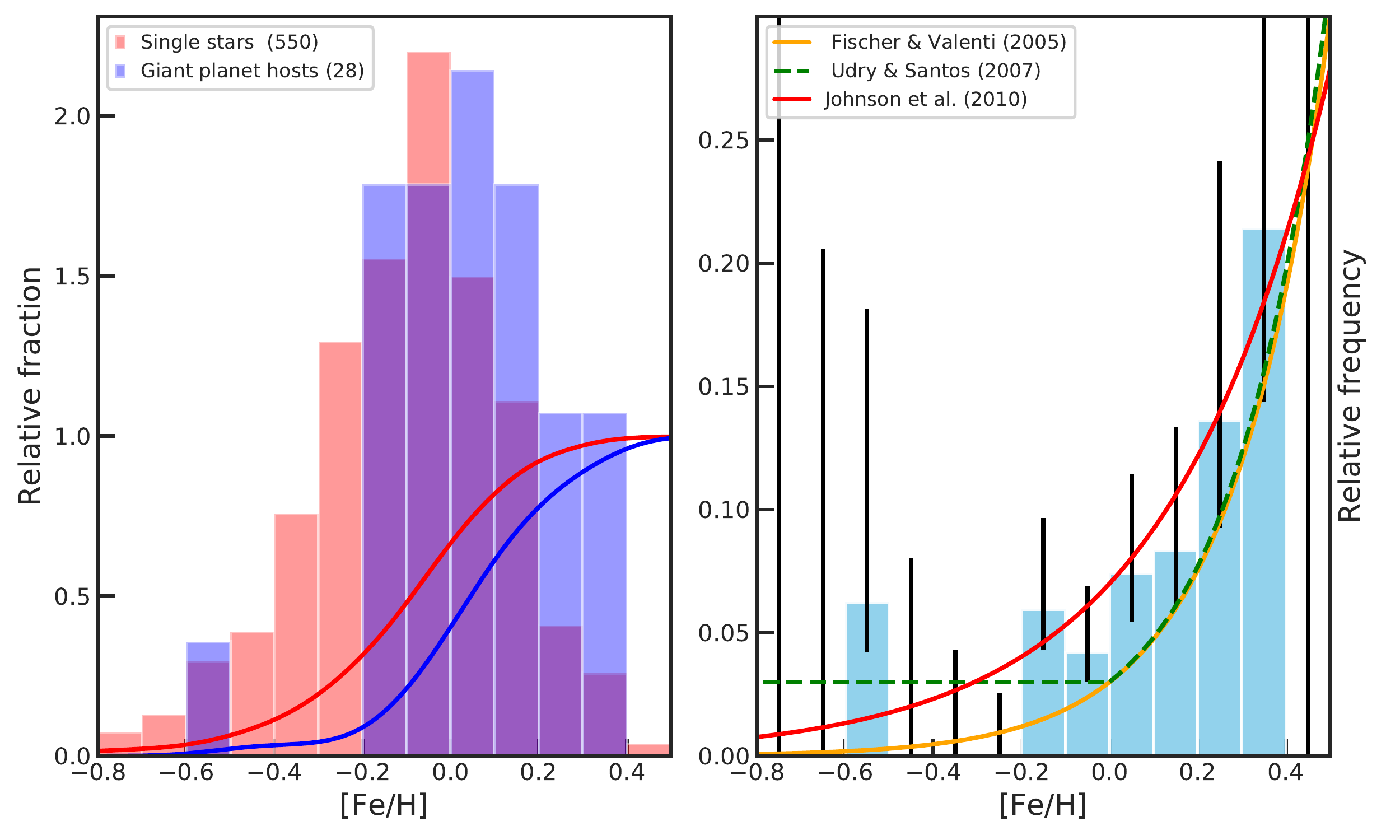}
\vspace{-0.2cm}
\caption{(Left panel): Metallicity distribution of stars hosting giant planets (blue)  and stars without any detected planets (red) from the HARPS planet search program. The KDE fit of the cumulative distribution of [Fe/H] for the two samples is also shown with the curves of corresponding colors. (Right panel): The dependence of giant planet frequency on stellar metallicity. The binomial errors of the planet fractions in each bin are estimated following \citet{cameron_estimation_2011}.}
\label{fig-giant_planet_metal_harps2}
\end{figure}

On the left panel of Fig.~\ref{fig-giant_planet_metal_harps2} I plot the metallicity distribution of SnoP and HMPH (again only planets 50~M$_{\oplus}$ $<$ M $<$ 13~M$_{\jupiter}$ are considered)  taken from the HARPS volume-limited (up to about 60 pc from the Sun) survey \citep{lo_curto_harps_2010}. This sample consist of 582 FGK stars, for which the stellar parameters, including the stellar metallicities, were homogeneously derived in \citet{sousa_spectroscopic_2011}. The figure shows that the mean metallicity of the HMPH sample (0.046$\pm$0.035) is again higher than the mean metallicity of the SnoP ($-$0.112$\pm$0.010). The performed KS test also suggest a significantly different metallicity distributions for the two samples (see Table~\ref{table_KS_metal_HMPH}).

\begin{table}[H] 
\caption{The results of the KS tests comparing the metallicity distributions of stars with and without planets. The p-values smaller than 0.05 are highlighted in boldface.} 
\centering
%% \tablesize{} %% You can specify the fontsize here, e.g.  \tablesize{\footnotesize}. If commented out \small will be used.

\begin{tabular}{p{6cm}p{2cm}p{2cm}}
\toprule
{Samples}	& {KS statistic}	& {KS p-value}\\
& & \\

\textbf{Figure~\ref{fig-giant_planet_metal}}	& 	& \\
\midrule
HMPH Transit vs HMPH RV		& 0.089			& 0.392 \\
HMPH RV vs SnoP		& 0.47			& \textbf{1.12e-40} \\
HMPH Transit vs SnoP		& 0.49			& \textbf{1.60e-30}\\
\end{tabular}

\bigskip

\begin{tabular}{p{6cm}p{2cm}p{2cm}}
\textbf{Figure~\ref{fig-giant_planet_metal_harps2}}	& 	& \\
\midrule
HMPH vs SnoP		& 0.339			& \textbf{0.003} \\
\end{tabular}

\bigskip

\begin{tabular}{p{6cm}p{2cm}p{2cm}}
\textbf{Figure~\ref{fig-very_giant_planets_homo}}	& 	& \\
\midrule
GPH vs SGMP		& 0.152			& \textbf{0.026} \\
GPH dwarfs vs SGMP dwarfs		& 0.067			& 0.935 \\
GPH giants vs SGMP giants		& 0.288			& 0.061 \\
GMP dwarfs vs SnoP dwarfs		& 0.450			& \textbf{1.92e-50}\\
GMP giants vs SnoP giants		& 0.281		& \textbf{0.007}\\
SGMP dwarfs vs SnoP dwarfs		& 0.480			& \textbf{2.45e-14}\\
SGMP giants vs SnoP giants		& 0.187			& 0.105\\
\end{tabular}

\bigskip

\begin{tabular}{p{6cm}p{2cm}p{2cm}}
\textbf{Figure~\ref{fig-low_mass_planet_metal}}	& 	& \\
\midrule
LMPH  vs SnoP		& 0.120			& 0.484 \\
Only LMPH  vs SnoP		& 0.252			& \textbf{2.06e-4}\\
\bottomrule
\end{tabular}

\label{table_KS_metal_HMPH}
\end{table}

The right panel of Fig.~\ref{fig-giant_planet_metal_harps2} shows the dependence of giant planet frequency (number of stars with planets relative to number of all surveyed stars) on stellar metallicity. The plot clearly shows that the frequency sharply increase with metallicty. If for sub-solar metallcities about 5\% of the stars have giant planets, at super-solar metallicities the relative frequency reaches to about 20\%. It is important to note that this giant-planet -- metallicity relation is practically unaffected by the possible dependence of  RV precision on [Fe/H]. In principle, metallicity affects the strength of spectral lines, and thus RV precision. However, for the stars with [Fe/H] $\gtrsim$  $-0.8$ dex, the RV precision is practically independent of metallicity\footnote{To my knowledge there is no study on the dependence of photometric precision on stellar metallicity for transiting planet surveys. However, the  detectability of transiting planets slightly depends on metallicity through the relationships between transit depth, stellar radius, and stellar metallicity: transiting planets are slightly easier to detect around metal-poor dwarf stars than around metal-rich ones \citep[][]{gaidos_objects_2013}.} \citep[][]{valenti_stellar_2008}.

In Fig.~\ref{fig-giant_planet_metal_harps2} (right panel) I also show the functional form of the planet occurrence -- metallicity relation derived in several works using different samples \citep{fischer_planet-metallicity_2005, udry_statistical_2007, johnson_giant_2010}. One can see that there is, at least, qualitative agreement between the fitted curves obtained by different authors for stars with metallicities above the solar value. These curves also describe well the increase of planet frequency for solar and super-solar metallicity stars observed for the HARPS sample in the figure. The main difficulty  and disagreement between the different works come for sub-solar metallicities. \citet{udry_statistical_2007}, based on the CORALIE sample from \citep{santos_spectroscopic_2004} and the sample from \citet{fischer_planet-metallicity_2005}, proposed that at sub-solar metallicities the giant planet frequency might be constant rather than exponential, which is clearly the case for solar and super-solar metallicities. This result has been contested by \citet{johnson_giant_2010} who used a larger sample of 1266 stars with and without planets from the California Planet Survey.  Based on a Bayesian analysis the authors found that giant planet occurrence rate is a strong exponential function of metallicity that continues to sub-solar metallicities. More recently, \citet{mortier_functional_2013} performed a statistical Bayesian analysis on large volume-limited samples of CORALIE, HARPS\footnote{This HARPS sample is the same as the one shown in Fig.~\ref{fig-giant_planet_metal_harps2}. The only difference is that a few planets have been discovered around these stars after the study of \citet{mortier_functional_2013}.}, and CORALIE+HARPS (the combined sample) to test between different functional forms of the planet--metallicity correlation. While confirming the general correlation between giant planet incident and metallicity, they  concluded that the samples are simply too small to be able to distinguish between a constant or an exponential form in the low-metallicity region. The authors pointed out that a sample of about 5000 stars (about a factor of 3 larger than the sample they used) is required to distinguish between different models (functional forms). Hopefully, this question will get its answer soon with the arrival of the ongoing and upcoming missions such as Gaia\footnote{The name "Gaia" was initially derived as an acronym for Global Astrometric Interferometer for Astrophysics.} \citep{sozzetti_detection_2001}, TESS \citep{ricker_transiting_2015}, and PLATO \citep{rauer_plato_2014}.

Using well defined samples one can determine the occurrence rate of given type of planets around given type of stars. Occurrence rate calculations require corrections for the detectability of a given type of planets. In case of  transit surveys in addition one should take into account the geometric probability for seeing a transit of a given type of planet. For a recent review on planet occurrence in RV and transit surveys I refer the reader to \citet{deeg_planet_2018}. The main properties of giant planets detected by RV and transit methods was recently reviewed by \citet{santerne_populations_2018}. The latter author compiled a list of occurrence rate estimations of giant planet  at different orbital periods: hot-Jupiters (P $<$ 10 days), period-valley giants (10 $<$ P $<$ 85 days), and temperate giants (85 $<$ P $<$ 400 days). Their results show that the average occurrence rate increases from about 1\% for the shortest period giants to about 3\% for giant planets with orbital periods of several hundred days \citep[see][and references therein]{santerne_populations_2018}. It is interesting to note that the occurrence rate calculations for hot-Jupiters detected by transit method are about factor of two lower than the one calculated for RV planets \citep[e.g.][]{santerne_sophie_2016}. The exact reason for this discrepancy is still unclear, but can be related to properties of the stars (e.g. binary rate) monitored in RV and transit surveys and/or (in)ability of accurate characterization of these stars \citep[e.g.][]{guo_metallicity_2017}. The integrated occurrence rate of giant planets (M $>$ 50~M$_{\oplus}$) with orbital periods shorter than 10 years is estimated to be 13.9$\pm$1.7\% \citep{mayor_harps_2011}.

\subsection{Metallicity of sub-Jupiters} \label{sub-Jupiters}

In Fig.~\ref{nayakshin_mordasini}, where I qualitatively compare the metallicities of different types of observed and synthetic planets, one can clearly see that there are two dips in the metallicity of observed planets at masses of $\sim$ 0.8~M$_{\jupiter}$ and 2~M$_{\jupiter}$. In this subsection I study these dips more in detail to understand whether they have an astrophysical origin or are due to different observational biases and selection effects.

\begin{figure}[H]
\centering
\includegraphics[width=15cm]{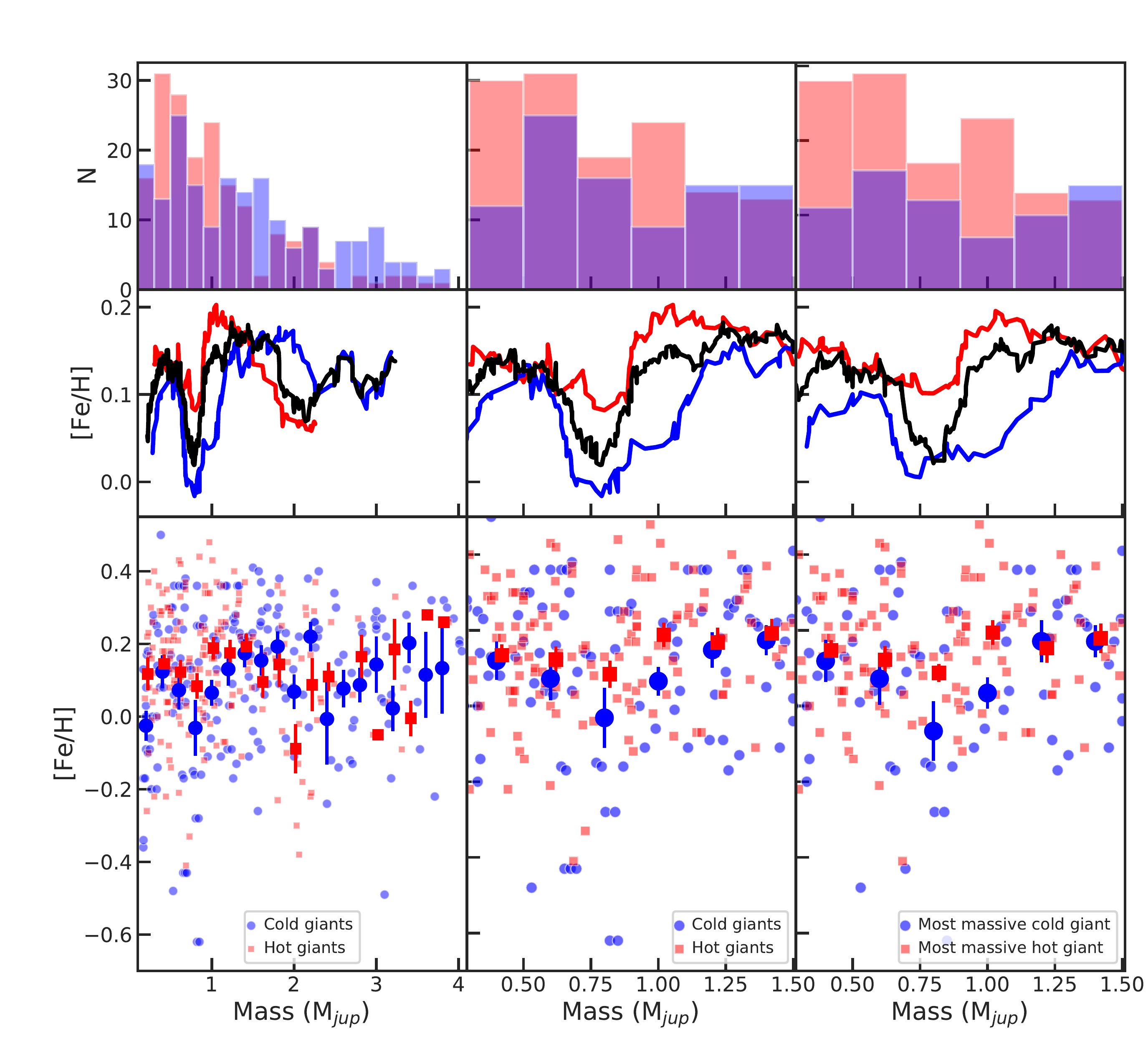}
\vspace{-0.2cm}
\caption{(Bottom panel): Dependence of the mass of hot giant planets on the metallicity of their host stars. The large symbols represent the mean metallicities and standard error of the mean (which does not include the measured uncertainty of [Fe/H] for individual points) for each mass bin (bin size is equal to 0.2~M$_{\jupiter}$).  The running mean of [Fe/H] as a function of mass of hot giants  (P $<$ 10 days; red), cold giants (P $>$ 100 days; blue), and all giant planets (in black) is shown in the middle panel. The running means were calculated using windows of 30, 30, and 40 for hot, cold, and all planets, respectively. The number of planets in each mass bin  is plotted on the top panels. In the middle column I show all the planets with masses between 0.3 and 1.5~M$_{\jupiter}$. In the leftmost column only the most massive planet in a system is considered.}
\label{fig-massive_planets_running_mean}
\end{figure}

In the leftmost panel of Fig.~\ref{fig-massive_planets_running_mean} I show the dependence of planet host metallicity on the mass of planets with masses between 50~M$_{\oplus}$ and 4~M$_{\jupiter}$. These planets are selected to be orbited around FGK dwarf stars (\logg \ $>$ 4.0 dex, 0.6~M$_{\odot} <$ M $<$ 1.5~M$_{\odot}$) with stellar parameters derived in a homogeneous way in SWEET-Cat. The running means of hot- (P $<$ 10 days) and cold-Jupiters (P $>$ 100 days)  with masses above about 1.5~M$_{\jupiter}$ have very different behaviors (see the middle panel). This is most probably due to small number of such planets in the sample. The small number of planets with masses above $\sim$ 1.5~M$_{\jupiter}$ and the change of the relative number of hot- and cold-Jupiters at 2~M$_{\jupiter}$ are probably responsible for the dip in metallicity observed at 2~M$_{\jupiter}$. The number of hot- and cold-Jupiters with masses between 0.3 and 1.5~M$_{\jupiter}$ (see the middle panel)  is relatively large and their relative fractions are quite balanced. This means that the metallicity dip observed at $\sim$ 0.8~M$_{\jupiter}$ might have a physical origin and deserves further exploration.

In the middle column of Fig.~\ref{fig-massive_planets_running_mean} one can see that both hot- and cold-Jupiters show a dip in metallicity at about $\sim$ 0.8~M$_{\jupiter}$, although the dip of cold Jupiters is more apparent. The main reason of separating hot and cold Jupiters is that orbital periods of giant planets correlate with the host star metallicity: long-period Jovians tend to orbit more metal-poor stars than hot-Jupiters \citep[e.g.][]{sozzetti_possible_2004, adibekyan_orbital_2013, narang_properties_2018, maldonado_chemical_2018}. This correlation can be seen in Fig.~\ref{fig-massive_planets_running_mean} as well, where the red squares and red curve lie above the blue circles and the blue curve.  To make the description of planets inside and outside of the metallicity dip easier and shorter, the planets inside the dip will be called 'sub-Jupiters', and the planets with masses higher and lower than the sub-Jupiters will be called 'Jupiters' and 'Saturns', respectively.

I tried to identify the lower and upper mass limits of sub-Jupiters by looking at a mass range for which the metallicity of their hosts stars is significantly lower than the metallicity of the'Jupiters' and 'Saturns'. Performed KS tests showed that the most significant difference between sub-Jupiters and non-sub-Jupiters is observed when sub-Jupiters are defined as planets with masses between about 200~M$_{\oplus}$ ($\sim$0.6~M$_{\jupiter}$) and 290~M$_{\oplus}$ ($\sim$0.9~M$_{\jupiter}$). However, when comparing the metallicities of the hosts of sub-Jupiters with those of the Saturns no statistically significant difference can be seen, while the difference from the hosts of Jupiters is always significant  (see Table~\ref{table_KS_sub_jupiters}). 

\begin{figure}[H]
\centering
\begin{tabular}{cc}
\includegraphics[width=7.5cm]{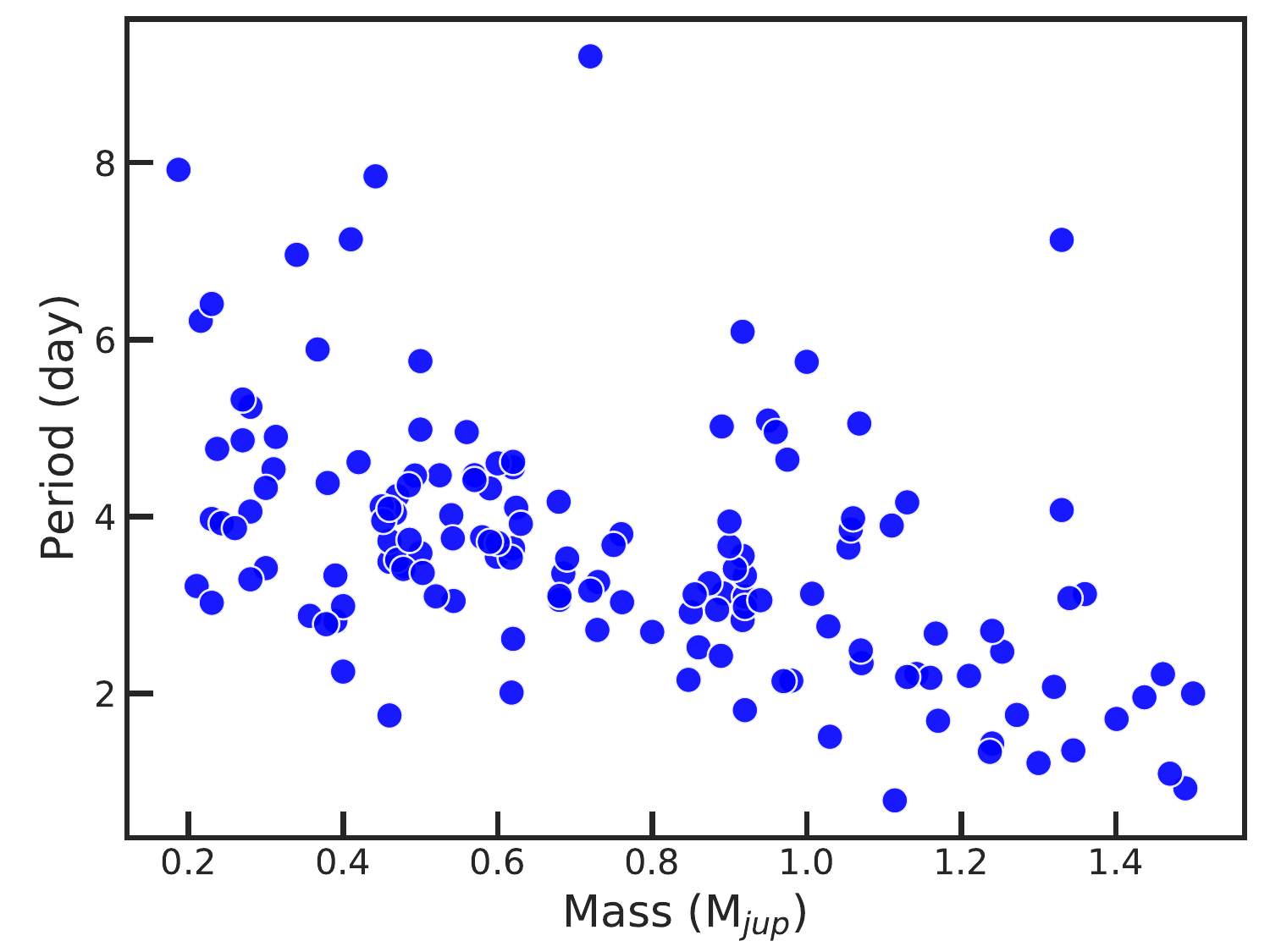}
\includegraphics[width=7.5cm]{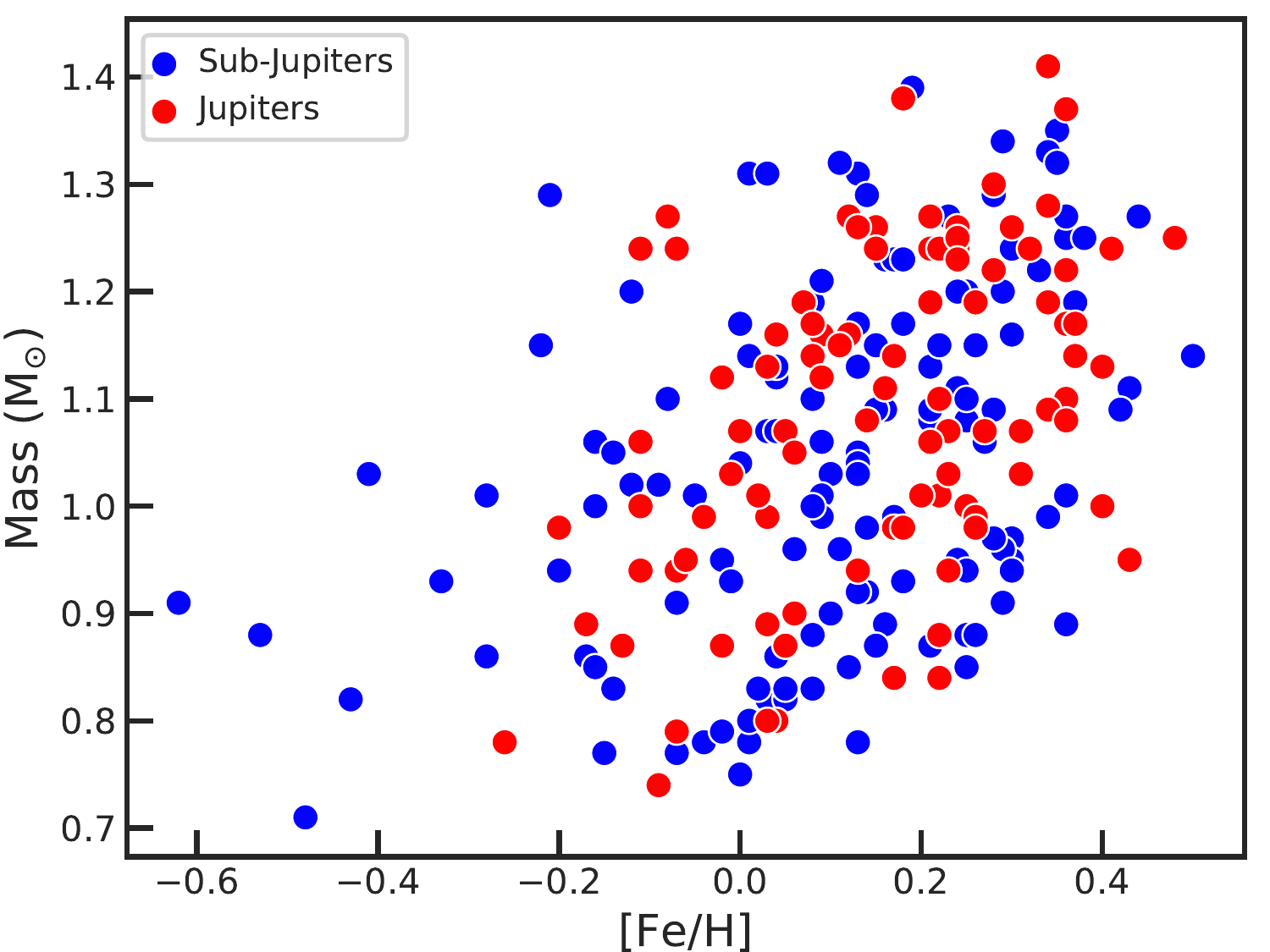}
\vspace{-0.4cm}
\end{tabular}
\caption{Orbital period against the mass of hot giant planets with 0.3 $<$~M$_{p}$ $<$ 1.5~M$_{\jupiter}$ (\textit{left panel}). In each planetary system, only the most massive planet is considered. Stellar mass against metallicity for sub-Jupiters and Jupiters (\textit{right panel}).}
\label{fig-subJupiters_period_mass}
\end{figure}

The next step was to investigate whether the sub-Jupiters (and/or their hosts) are different from the Jupiters and Saturns (and/or their hosts) in parameters that are linked to planet formation processes (e.g. orbital period, eccentricity, multiplicity, host star mass). To test whether the obtained results are sensitive to the multiplicity of the planets, I also considered the most massive planet for each planetary system (see rightmost panel of Fig.~\ref{fig-massive_planets_running_mean} and Table~\ref{table_KS_sub_jupiters}). The results did not differ significantly from the case where all the planets in each planetary system were considered.  The most significant difference was observed for orbital periods of hot giant planets (Table~\ref{table_KS_sub_jupiters}). In the left panel of Fig.~\ref{fig-subJupiters_period_mass} the dependence of orbital periods on the mass of the sample of hot giant planets is plotted. The observed clear correlation is the cause of the different orbital periods observed for planets with different masses. The lack of very short period (P $<$ 2 days) sub-Jupiters probably cannot be explained by photoevaportation as these planets loose very small fraction of their mass \citep[e.g.][]{murray-clay_atmospheric_2009, adams_magnetically_2011, owen_planetary_2012}.  Worth mentioning that this process is invoked to explain the lack of super-Earths and Neptunes at very short orbits \citep[e.g.][]{owen_kepler_2013, lundkvist_hot_2016, mazeh_dearth_2016}. Most of the mechanisms proposed to explain the orbital distribution  and the observed lower boundary of hot-Jupiters include disk migration \citep[e.g.][]{nelson_evidence_2017, dawson_origins_2018} and high-eccentricity migration \citep[e.g.][]{matsakos_origin_2016, nelson_evidence_2017, owen_photoevaporation_2018}, both followed by tidal dissipation. Very recently, \citet{bailey_hot_2018} proposed that the lower boundary of the periods of hot-Jupiters can be a result of their \textit{in-situ} formation followed by tidal decay for the most massive planets. Whatever is the mechanism shaping the orbital architecture of the hot-Jupiters, it is not straightforward to link it with the significant  metallicitiy differences observed between the sub-Jupiter and Jupiter host stars.

\begin{table}[H] 
\caption{The results of the KS tests comparing the orbital periods and host stellar masses of giant planets of different masses. The number of planets in each sub-sample is presented in parenthesis. The p-values smaller than 0.05 are highlighted in boldface.}
\centering
%\tablesize{\footnotesize}
%% \tablesize{} %% You can specify the fontsize here, e.g.  \tablesize{\footnotesize}. If commented out \small will be used.

\begin{tabular}{p{6.5cm}p{2.5cm}p{2cm}p{2cm}}
\toprule
{Samples} & {Parameter}	& {KS statistic}	& {KS p-value}\\

\textbf{Cold planets (P $>$ 100 days)}	& 	& & \\
\midrule
Sub-Jupiters (28) vs Saturns (25)	& [Fe/H]	& 0.34			&  0.061 \\
Sub-Jupiters (28) vs Saturns (25)	& Period	& 0.24			& 0.366 \\
Sub-Jupiters (28) vs Jupiters (44) & [Fe/H]	& 0.38		&  \textbf{0.009} \\
Sub-Jupiters (28) vs Jupiters (44) & Period	& 0.24			&  0.240\\
\end{tabular}
\smallskip

\begin{tabular}{p{6.5cm}p{2.5cm}p{2cm}p{2cm}}
\textbf{Most massive cold planets (P $>$ 100 days)}	& 	& & \\
\midrule
Sub-Jupiters (21) vs Saturns (18)  & [Fe/H]	& 0.31			&  0.231 \\
Sub-Jupiters (21) vs Saturns (18)  & Period	& 0.43			& \textbf{0.034} \\
Sub-Jupiters (21) vs Jupiters (35) & [Fe/H]	& 0.37		&  \textbf{0.039} \\
Sub-Jupiters (21) vs Jupiters (35) & Period	& 0.26			& 0.260 \\
\end{tabular}
\smallskip

\begin{tabular}{p{6.5cm}p{2.5cm}p{2cm}p{2cm}}
\textbf{Hot planets (P $<$ 10 days)}	&  &	& \\
\midrule
Sub-Jupiters (26) vs Saturns (51)	& [Fe/H]	& 0.22		&  0.332 \\
Sub-Jupiters (26) vs Saturns (51)	& Period	& 0.39			&  \textbf{0.005}\\
Sub-Jupiters (26) vs Jupiters (50) & [Fe/H]	& 	0.34		&  \textbf{0.023} \\
Sub-Jupiters (26) vs Jupiters (50) & Period	& 	0.42		&  \textbf{0.002}\\
\end{tabular}
\smallskip

\begin{tabular}{p{6.5cm}p{2.5cm}p{2cm}p{2cm}}
\textbf{Most massive hot planets (P $<$ 10 days)}	&  &	& \\
\midrule
Sub-Jupiters (24) vs Saturns (47)	 & [Fe/H]	& 0.28		&  0.134 \\
Sub-Jupiters (24) vs Saturns (47) & Period	& 0.37			& \textbf{0.018} \\
Sub-Jupiters (24) vs Jupiters (47) & [Fe/H]	& 0.36			& \textbf{0.021} \\
Sub-Jupiters (24) vs Jupiters (47) & Period	& 0.45			& \textbf{0.002} \\
\bottomrule
\end{tabular}

\label{table_KS_sub_jupiters}
\end{table}

Besides the metallicity, other characteristics of the protoplanetary disks, such as disk lifetime, disk accretion rate and the disk mass, have very strong influence on the formation and evolution of giant planets \citep[e.g.][]{mordasini_extrasolar_2012, hasegawa_planetary_2013, hasegawa_planet_2014}. In fact all these parameters strongly inter-correlate \citep[e.g.][]{mordasini_extrasolar_2012, hasegawa_planetary_2013}. While no direct information is available about the protoplanetary disk properties where the observed planets have been formed, as a proxy one can consider the stellar mass. The disk mass linearly \citep[e.g.][]{andrews_mass_2013, mohanty_protoplanetary_2013} or superlinearly \citep[M$_{disk}$ $\propto$ (M$_{star})^{>1}$;  e.g.][]{barenfeld_alma_2016, ansdell_alma_2016, pascucci_steeper_2016} correlates with the mass of the host star with an average disk-to-star mass ratio of $\sim$0.2\%--0.6\% \citep[e.g.][]{andrews_mass_2013}. Thus massive stars holding massive disks should have higher probability to form giant planets \citep[e.g.][]{kennedy_planet_2008}. However, these massive disks disperse  earlier \citep[e.g.][]{ribas_protoplanetary_2015}, which makes the formation of massive cores of gas giant planets difficult. At the same time, the migration of giant planets formed in these short-lived disks is supposed to be very limited \citep[][]{burkert_separation/period_2007, currie_semimajor_2009}. To make the picture more complex one should remember that the disk mass  also correlates with the disk accretion rate \citep{manara_evidence_2016}, a parameter which might be important for the fast formation of the cores of giant planets \citep{liu_migration_2016}.

Interestingly, no significant difference of stellar mass was found for the subsamples of giant planets suggesting that for a given stellar mass Jupiters tend to form around more metallic stars than their lower mass counterparts. This is somehow surprising since the stellar mass correlates with the stellar metallicity as shown in the right panel of Fig.~\ref{fig-subJupiters_period_mass}. This correlation is a consequence of stellar evolution, stellar age -- metallicity relation, and the typical color and magnitude cuts used in the selection of target in planet search programs \citep[][]{santos_statistical_2003, fischer_planet-metallicity_2005, ghezzi_stellar_2010, johnson_giant_2010}. With such results at hand, one might argue that the higher metallicity of Jupiter hosts would mean longer disk lifetime \citep{ercolano_metallicity_2010} and higher amount of planet building material \citep[e.g.][]{mordasini_extrasolar_2012}. These two parameters can directly affect the formation efficiency of giant planets and their migration rate. For cold giant planets no significant difference in orbital periods between sub-Jupiters and Jupiters is observed. This might be due to difference in starting positions of planet formation in disks with different metallicities \citep[e.g.][]{mordasini_extrasolar_2012}.

While in this section a very qualitative and somewhat speculative discussion was developed to explain the low metallicity of sub-Jupiter hosts when compared to the stars hosting Jupiter-mass planets, a more quantitative analysis is necessary to understand the origin of this metallicity difference. Perhaps, even before such an analysis it is necessary to study the influence of different selection and observational biases on this finding.

\subsection{Very massive giants and metallicity} \label{super-giants}

In this subsection I will discuss the dependence of very massive planet formation on stellar metallicity. However, such a discussion requires a knowledge of the upper mass boundary of planets, which is still an open question. As mentioned in Sect.~\ref{giant_planet_metal}, depending on the goal of the study, different authors use different lower and upper mass limits when defining giant planets. 13~M$_{\jupiter}$ is usually used as an upper mass limit for giant planets  to separate from sub-stellar mass objects with significant central deuterium burning \citep[e.g.][]{spiegel_deuterium-burning_2011}. However, as already mentioned in the manuscript, the exact deuterium burning mass-limit depends on several parameters of the object. Another possible limitation of mass-limit based definition is that it does not involve  formation mechanism of the object. Finally, the mass distribution of sub-stellar objects show no any special feature around this mass limit \citep[e.g.][]{udry_detection_2010}. \citet{schlaufman_evidence_2018} proposed that the separation between giant planets and BDs should be somewhere between 4~M$_{\jupiter}$ to 10~M$_{\jupiter}$, assuming that the sub-stellar objects formed via CA and GI can be statistically well separated. In particular, \citet{schlaufman_evidence_2018} assumed that the giant planets preference for  metal-rich host stars is an indication of their formation through CA \citep[but see][]{nayakshin_dawes_2017}.  \citet{schneider_defining_2011}, in exoplanet.eu set an upper limit of 25~M$_{\jupiter}$ for giant planets. Although appreciating the importance of having a formation-based definition for planets, the authors correctly noted about the difficulty (or at this point even probably impossibility) of knowledge about the formation mechanism for the planets.  The choice of 25~M$_{\jupiter}$  the authors justified by the start of a dip at around this mass in the observed distribution in $M\sin i$ of substellar mass objects \citep[e.g.][]{mayor_hot_2005, grether_how_2006, udry_detection_2010, sahlmann_possible_2011}.

Exoplanets are sometimes categorized into different populations defined by distinct metallicity distributions \citep[e.g.][]{buchhave_three_2014}. \citet{ribas_eccentricity-mass_2007} found that the average metallicity of stars with planets of mass $<$ 4~M$_{\jupiter}$ is lower than that of higher mass planet hosts by about 0.15 dex. However, in their comparison the authors did not separate the super-Earth and Neptune mass planets from giant planets. Moreover, these low mass planets in their sample mostly had super-solar metallicities, while the majority of later-on detected low-mass planets hosts have subsolar or solar metallicities \citep[e.g.][]{mayor_harps_2011, buchhave_abundance_2012} if not orbiting their stars at very short periods \citep{adibekyan_orbital_2013, mulders_super-solar_2016}. Using a significantly larger sample of giant planets with masses greater than 0.1~M$_{\jupiter}$ and with homogeneously derived host metallicities derived in SWEET-Cat, \citet{adibekyan_orbital_2013} revisited the metallicity distribution of very massive giants. The authors found that the stars hosting massive Jupiters (mass $>$ 4~M$_{\jupiter}$) have an average metallicity of 0.083 $\pm$ 0.032 dex, while the stars hosting planets with masses between 1 and 4~M$_{\jupiter}$ had an average metallicity of 0.149 $\pm$ 0.016 dex. Despite the lower average metallicity (lower by 0.066 $\pm$ 0.036 dex) of massive Jupiter hosts when compared to the relatively lower mass Jupiter (1 $<$ M$_{p}$ $<$ 4~M$_{\jupiter}$) hosts, the performed KS statistics suggested  a probability of about 15\%  that the two subsamples have the same underlying metallicity distribution. \citet{adibekyan_orbital_2013} concluded that a simple separation in mass at 4~M$_{\jupiter}$ does not reveal two different populations in metallicity.

Recently, \citet{santos_observational_2017} using the SWEET-Cat database also addressed the question whether very massive Jupiters tend to form at metallicities lower than their less massive counterparts. The authors found a convincing observational evidence that giant planets with masses above and below $\sim$4~M$_{\jupiter}$ constitute two distinct populations\footnote{Recently, when studying mass-radius relation for exoplanets, \citet{bashi_two_2017} also found an evidence of existence of a transition at a planetary mass  of $\sim$ 5~M$_{\jupiter}$.}. They showed that the hosts of giant planets with masses $<$ 4~M$_{\jupiter}$ have metallicities on average higher than that of the fields stars without planets. At the same time, the hosts of their most massive planets were more massive stars with relatively low metallicities, similar to that observed in field stars of similar mass. Compensation of the lack of metallicity by higher stellar mass (thus high disk mass and availability of larger amount of planet building material) was not considered as a convincing explanation. Based on the aforementioned observational results \citet{santos_observational_2017} proposed that the planets in the two mass regimes may form in different ways: low-mass giant planets are formed in metal-rich disks through CA \citep[e.g.][]{mordasini_extrasolar_2012} and the more massive planets are formed in massive and metal-poor disks via GI \citep[e.g.][]{rafikov_can_2005, cai_effects_2006}.

It is interesting to see that the transiting planets with a radii larger than about 10 R$_{\oplus}$ (note that the radius of Jupiter is 11.2 R$_{\oplus}$) also seem to have an average metallicity slightly (perhaps statistically not significant) lower than the giant planets with smaller radii \citep[see Fig.3, Fig. 7, and Fig. 6 of][ respectively]{petigura_california-kepler_2018, narang_properties_2018, wang_giant_2018}. However, the number of transiting planets with very large radii and precise host metallicities is small and it is difficult to conclude whether there is a break-point radius above which planet hosts show low metallicities. Perhaps, besides the sample size and precision in metallicity, the main limitation of testing the aforementioned hypothesis is the very weak correlation between mass and radius for these H/He dominated giant planets \citep[e.g.][]{weiss_mass_2013, hatzes_definition_2015, chen_probabilistic_2017}.

Very recently, \citet{schlaufman_evidence_2018} also studied the upper mass limit of exoplanets in connection with metallicity distributions of planets and sub-stellar mass objects. The author used a sample of 119 systems with planetary-mass companion having both RV and transit signals and true masses\footnote{Observations of the transits ensures that the inclination is about 90$^\circ$ and thus $\sin i \approx 1$.} $>$ 0.1~M$_{\jupiter}$ . These planets had orbits mostly within 10 days. His sample of subs-stellar objects consisted of 27 objects with masses $<$ 300~M$_{\jupiter}$. The constrain of having RV and transit signal restricted the orbital periods of the subs-stellar mass objects to be mostly within 50 days. First, \citet{schlaufman_evidence_2018} applied a clustering algorithm to separate the giant planets and non-planets based on the metallicities of their host/primary stars. The objects with M $<$ 4~M$_{\jupiter}$ and M $>$ 10~M$_{\jupiter}$ the algorithm unanimously classified as belonging to two different groups: giant planets and BD/low-mass stars, respectively.  The clustering analysis of \citet{schlaufman_evidence_2018} using the SWEET-Cat sample suggested a separation at a similar mass of 4 ~M$_{\jupiter}$ (note the similarity in the separation mass proposed by \citet{santos_observational_2017}). In the second part of the analysis, ``Moving Median Analysis'', the author estimated the  mass that separates the giant planets that are formed exclusively through CA and objects that do not show preference towards the high metallicity of their hosts and thus, cannot be formed through CA. Besides the assumption that the giant planets form only through CA, \citet{schlaufman_evidence_2018} also build the analysis on the theoretical prediction of \citet{mordasini_extrasolar_2012} that ``massive objects formed by core accretion should only occur around the most metal-rich primaries''. He found that the point at which secondaries do not preferentially orbit metal-rich primary stars occurs at about 10~M$_{\jupiter}$. 

Inclusion of stellar mass binary systems in the sample of \citet{schlaufman_evidence_2018} might decrease the average metallicity of the massive secondaries sample, which in turn, would impact on the estimation of the breakpoint mass separating the giant planets from BDs.  Recent observations suggest that the close (period $<$ 100 - 10,000 days, depending on the work) binary fraction of solar type stars strongly anti-correlates with metallicity \citep{raghavan_survey_2010, gao_binarity_2014, moe_close_2018}. In particular, \citet{moe_close_2018} showed that the binary fraction decreases from 40\% to 10\% in the mtallicity range of $-$1.0 dex and 0.5 dex. It is interesting to note that the wide orbit binary fraction of solar-type stars and  close binary fraction of O- and B-type stars does not significantly depend on the metallicity across $-$1.5 $<$ [Fe/H] $<$ 0.5 dex range.

\begin{figure}[H]
\centering
\includegraphics[width=14cm]{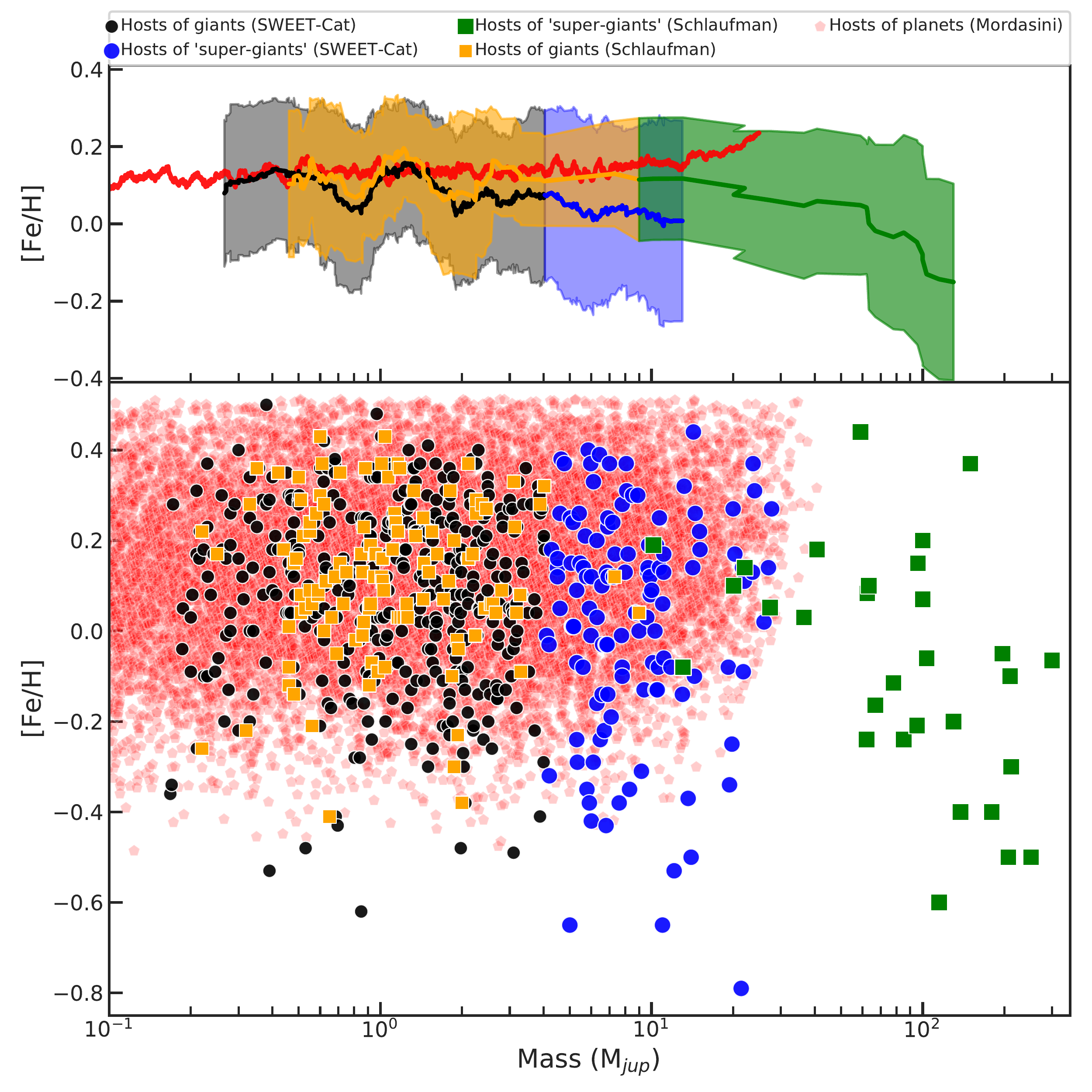}
\vspace{-0.2cm}
\caption{(\textit{Bottom panel}) Stellar metallicity against mass of giant planets and super-giant planets/sub-stellar objects discussed in \citet{schlaufman_evidence_2018} (yellow and green) and  available in the SWEET-Cat database (black and blue). The synthetic planets predicted by the CA model of \citet{mordasini_extrasolar_2012} (red) are also plotted for a visual comparison. (\textit{Top panel}) The running mean of [Fe/H] as a function of mass of the aforementioned objects. The running means were calculated using windows of 20, 50, and 200 for the \citet{schlaufman_evidence_2018}, SWEET-Cat, and CA \citep{mordasini_extrasolar_2012} planets, respectively. These numbers reflect the sizes of each sample. For the \citet{schlaufman_evidence_2018} and SWEET-Cat samples  the 1$\sigma$ uncertainty in the moving mean is also shown as a yellow-green and black-blue regions, respectively.}
\label{fig-schlaufman_mordasini_planet_metallicity}
\end{figure}

In the bottom panel of Fig.~\ref{fig-schlaufman_mordasini_planet_metallicity} I show the dependence of mass of giant planets (yellow) and sub-stellar objects (green) on the metallicity of the host star from the sample of \citet{schlaufman_evidence_2018}. In the same figure I also plot the giant planets (masses between 50~M$_{\oplus}$ and 4~M$_{\jupiter}$) and super-giant planets (masses greater than 4~M$_{\jupiter}$) orbiting around stars with metallicities homogeneously derived in SWEET-Cat. I  note that while the masses of the objects from the sample of \citet{schlaufman_evidence_2018} are true masses, the masses for the SWEET-Cat objects are mostly the minimum masses ($M\sin i$). For a comparison, I also show the CA predicted synthetic planets of \citet{mordasini_extrasolar_2012}. The figure clearly shows that most of the giant planets and super-giant planets/BD with masses bellow about 20~M$_{\jupiter}$ share the area occupied by the synthetic CA planets. This means that, in principle, CA model of \citet{mordasini_extrasolar_2012} can explain the most metallic super-giant planets/BDs. The figure also shows that the area of M $\gtrsim$ 4~M$_{\jupiter}$ and [Fe/H] $\lesssim -$ 0.1 dex is very sparsely populated by the CA planets and the formation of the observed super-giant planets and/BDs in this area cannot be easily explained with the CA model of \citet{mordasini_extrasolar_2012}: at least not with the parameters (e.g. disk lifetime, disk mass) that were adopted in that specific model. It is interesting to see that almost all the sub-stellar objects with masses below about 60-70~M$_{\jupiter}$ of the sample of \citet{schlaufman_evidence_2018} are observed at super-solar metallicities. At the same time, almost all the low-mass stars\footnote{The hydrogen-burning mass limit for stars is at about 85~M$_{\jupiter}$ depending on the metallicity \citep[e.g.][]{hayashi_evolution_1963, baraffe_evolutionary_1998, nakano_pre-main_2014}.} with masses above about 150~M$_{\jupiter}$ have sub-solar metallicities. This latter trend is expected as the stellar binary fraction decreases with metallicity \citep[e.g.][]{moe_close_2018}. It is also interesting to see (compare the yellow and black/blue curves on the top panel of Fig.~\ref{fig-schlaufman_mordasini_planet_metallicity}) that the giant planets of \citet{schlaufman_evidence_2018} are on average more metallic (0.119 $\pm$ 0.015 dex) when compared with the giant planets from SWEET-Cat (0.091 $\pm$ 0.009 dex). This difference is probably due to the different orbital period distributions of the two samples. As discussed in Sect.~\ref{sub-Jupiters}, the short period Jupiters tend to orbit around more metallic stars than their longer period counterparts \citep[e.g.][]{ adibekyan_orbital_2013, maldonado_chemical_2018}.

\begin{figure}[H]
\centering
\includegraphics[width=12cm]{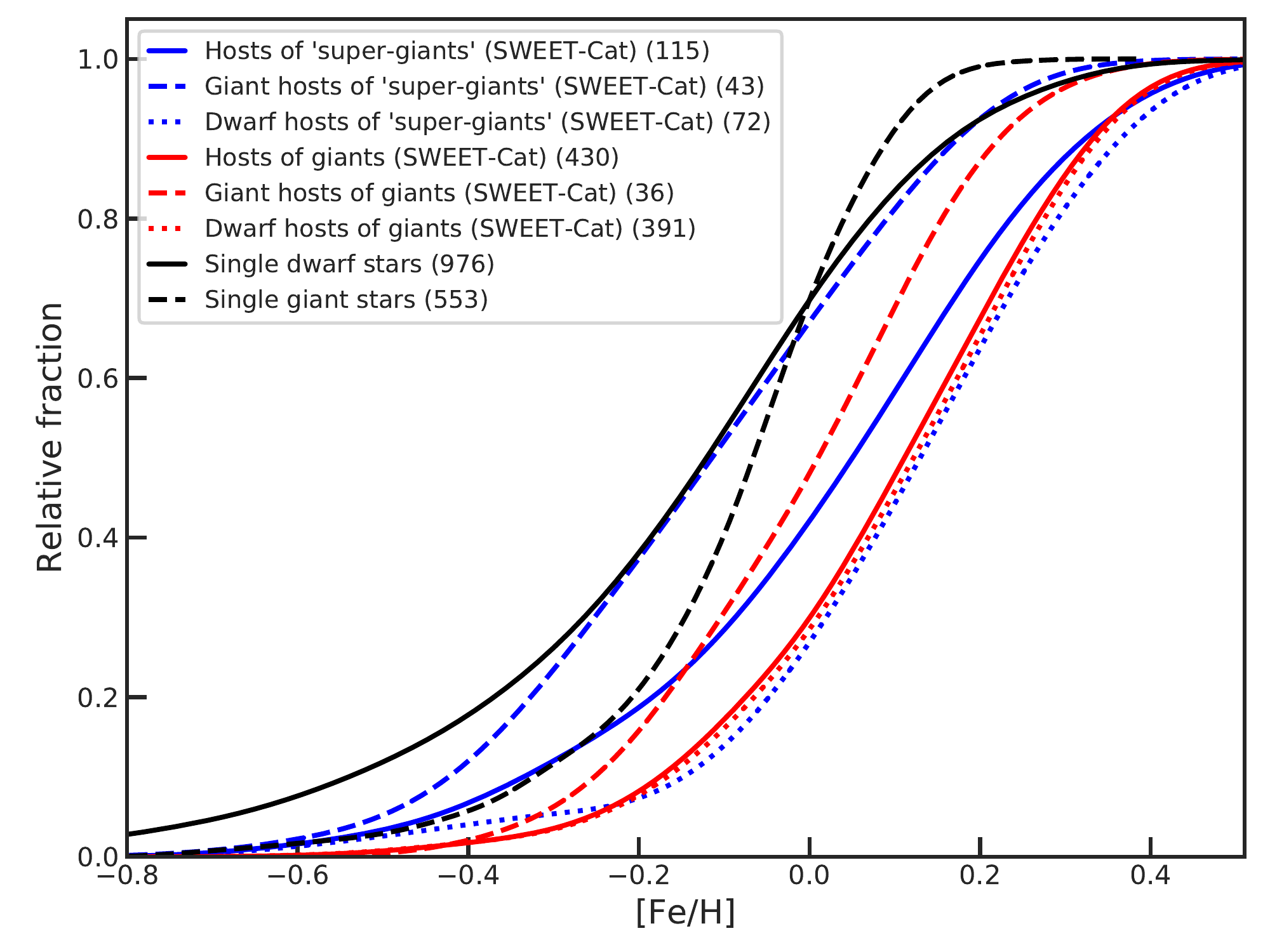}
\vspace{-0.2cm}
\caption{The KDE fit of the cumulative distributions of [Fe/H] for stars hosting giant (red) and super giant planets/BDs  (blue) together with the samples of solar neighborhood single dwarf and giant stars are shown in curves of different colors that are indicated in the plot. The planet hosting dwarf stars with masses less than 1.5~M$_{\odot}$ and giant stars with masses greater than 1.5~M$_{\odot}$ are shown with dotted and dashed curves, respectively.}
\label{fig-very_giant_planets_homo}
\end{figure}

Fig.~\ref{fig-very_giant_planets_homo} shows the metallicity distributions of stars hosting giant planets (GPH) with masses 50~M$_{\oplus}$ $<$ M $<$ 4~M$_{\jupiter}$ and super-giant planet (SGPH) with masses 4~M$_{\jupiter}$ $<$ M $<$ 20~M$_{\jupiter}$. The figure and performed KS statistics suggest that the giant planet hosts are significantly more metallic (see Table~\ref{table_KS_metal_HMPH}) than the hosts of more massive planets. The average metallicity of GPH is 0.091 $\pm$ 0.009 dex and the average metallicity of SGPH is 0.021 $\pm$ 0.022 dex. Despite the significant differences in the average metallicity distributions of giant planets and super-massive planets and/or BDs one probably should not make a bold conclusion about their different formation channels. These planets might have formed via the same mechanisms in different environments. As discussed in \citet{santos_observational_2017}, the hosts of super-giant planets on average are more massive than the hosts of Jupiter-like planets which means that the super-giant planets have been formed in more massive disks. 

\begin{figure}[H]
\centering
\includegraphics[width=14cm]{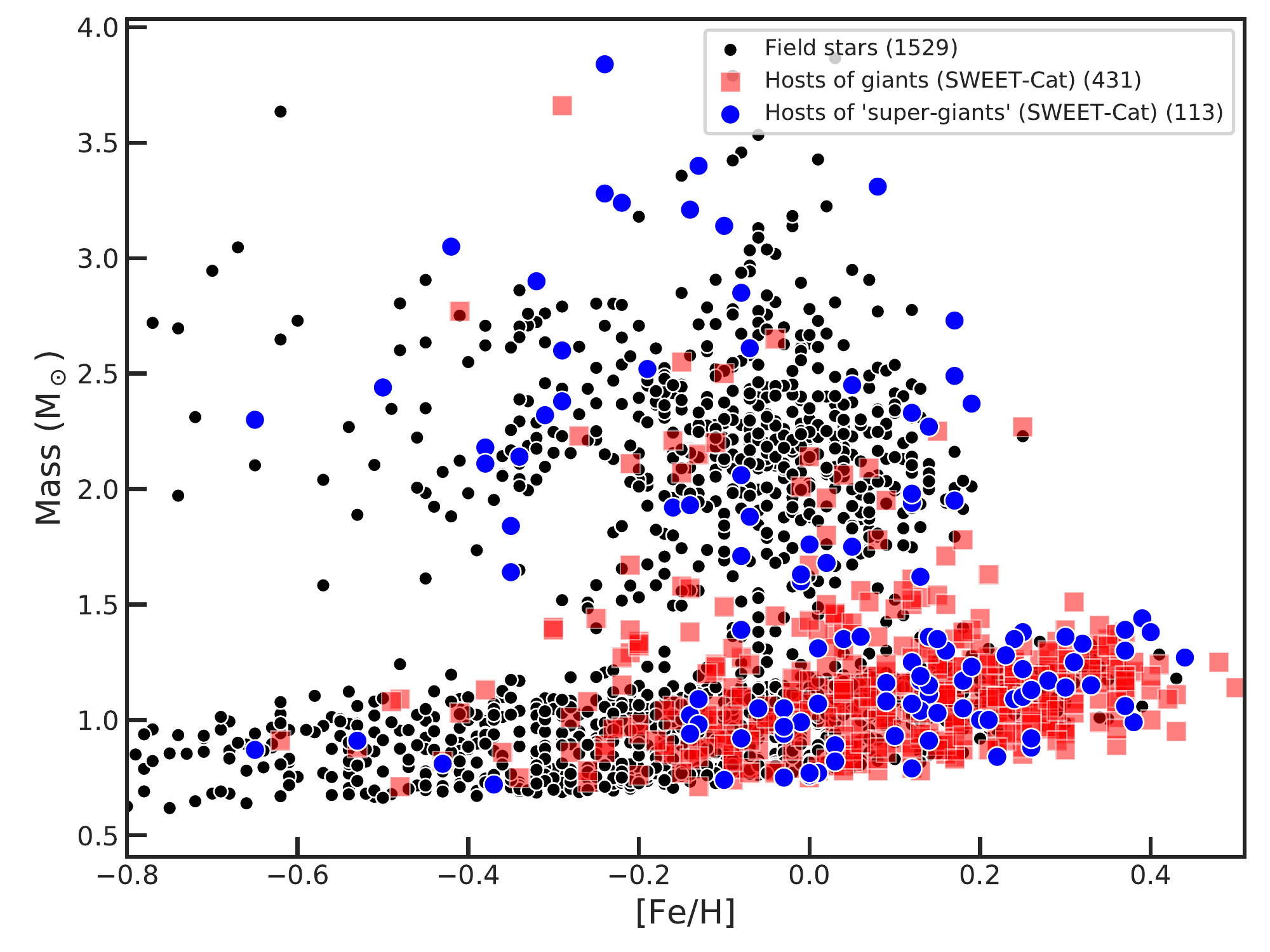}
\vspace{-0.2cm}
\caption{Dependence of stellar mass on [Fe/H] for stars hosting giant (red) and super giant planets/BDs  (blue) together with the samples of solar neighborhood single dwarf and giant stars (black).}
\label{fig-mass_metallicity}
\end{figure}

In order to study the impact of stellar mass on the obtained results I divide the samples into dwarf host stars (M $<$ 1.5~M$_{\odot}$)  and giant host stars (M $>$ 1.5~M$_{\odot}$). The stellar masses were derived using the calibration presented in \citet{torres_accurate_2010}. This calibration is based on stellar temperature, surface gravity, and metallicity of the stars. Fig.~\ref{fig-very_giant_planets_homo} and Table~\ref{table_KS_metal_HMPH}  show that when considering only dwarf stars the metallicity distribution of GPH and SGPH are almost identical. The average metallicity of SGP dwarf hosts is 0.106 $\pm$ 0.024 dex and the average metallicity of GP dwarf hosts is 0.100 $\pm$ 0.009 dex. Almost all the dwarf stars hosting super-giant planets have solar and super-solar metallicities (see Fig.~\ref{fig-mass_metallicity}). Such a metallicity distribution is expected for planets formed through CA \citep[e.g.][]{mordasini_extrasolar_2012} rather than GI/TD \citep{nayakshin_tidal_2015-2, nayakshin_dawes_2017}. Only four dwarf stars (HD111232, HD114762, HD181720, and HD22781) hosting SGP have  metallicities bellow $-$0.3 dex. It is worth noting that most of the dwarf stars hosting planets with [Fe/H] $<$ $-$0.3 dex are enhanced in $\alpha$-elements, such as Mg and Si \citep{haywood_peculiarity_2008, adibekyan_exploring_2012, adibekyan_overabundance_2012}. $\alpha$-enhanced iron-poor stars are also enhanced in oxygen \citep[e.g.][]{bertran_de_lis_oxygen_2015} and carbon \citep[e.g.][]{delgado_mena_chemical_2010}. Enhancement in such abundant elements as O, C, Mg, and Si makes these stars not very metal-poor, but iron-poor. Interestingly, the picture is drastically different when comparing the metallicity distributions of SGP and GP giant hosts. The average metallicity of SGP giant hosts is $-$0.122 $\pm$ 0.031 dex and the average metallicity of GP giant hosts is $-$0.005 $\pm$ 0.027 dex. The performed KS test suggest a probability of 0.06 that the metallicies of the two samples come from the same parent distribution. In general, the data shows that giant stars hosting planets (giant and super-giant) are less metallic than the dwarf hosts and that practically there are no giant stars hosting planets with super solar metallicities [Fe/H] $>$ 0.2 dex (see Fig.~\ref{fig-mass_metallicity}).

To  understand better the reason of dwarf and giant stars hosting planets having different metallicity distributions, in Fig.~\ref{fig-mass_metallicity} I show the distribution of stars with and without planets in the stellar mass -- metallicity diagram. It is important to note that the stars without planets have been searched for planets. The FGK field dwarf stars without detected planets have been observed within the HARPS GTO planet search program \citep{adibekyan_chemical_2012} and GK field giant stars without planets have been observed as part of the CORALIE planet search program \citep{alves_determination_2015} and the Okayama Planet search program  \citep{takeda_stellar_2008}. The masses of these field stars were derived using the same calibration formulae \citep{torres_accurate_2010} as for the exoplanet host stars. Fig.~\ref{fig-mass_metallicity} clearly shows that all the giant stars with and without planets have a limiting maximum metallicity of about 0.2 dex and the fraction of very low metallicity (e.g. [Fe/H] $<$ -0.4 dex) giant stars is much less when compared with the lower mass dwarf stars in the same metallicity region. Several studies have already observed this tendency of evolved, giant stars lacking the metal-rich and very metal-poor stars \citep[e.g.][]{taylor_widths_2005,luck_giants_2007,takeda_stellar_2008, ghezzi_metallicities_2010, adibekyan_chemical_2015}. Due to their shorter main sequence lifetimes, most of these giant and evolved (the stars with M $>$ 1.5~M$_{\odot}$ have surface gravities between 1.5 and 3.0 dex) stars are younger than their dwarf counterparts. The younger age together with the age--metallicity dispersion relation \citep[e.g.][]{da_silva_basic_2006, casagrande_new_2011}  might explain the narrower metallicity distribution of the
giant stars. These young stars are mostly local since they do not have time to migrate within the Galaxy \citep[e.g.][]{wang_influence_2013, minchev_chemodynamical_2013}. If the lifetime of the stars is long, then the radial migration in the disc would bring them from the inner,
metal-rich Galaxy \citep[e.g.][]{wang_influence_2013, minchev_chemodynamical_2013}. In addition, it is important to note that most large programs to search for planets around giant stars make a sample selection based on  a cut-off in the B -- V colour. This B -- V cut-off result in the lack of low-gravity and massive stars with high-metallicities \citep[][]{mortier_new_2013}. 

Besides comparing the metallicitis of the hosts of massive and very massive planets, it is very informative to compare their metallicity distributions  with the distribution of stars with not detected planets. The cumulative distribution of [Fe/H] for FGK field dwarf stars of \citet{adibekyan_chemical_2012}, and GK field giant stars of \citet{alves_determination_2015} and \citet{takeda_stellar_2008} is shown in Fig.~\ref{fig-very_giant_planets_homo}. The figure and the performed KS statistics (see Table~\ref{table_KS_metal_HMPH}) suggest that the dwarf hosts of both GP and SGP are significantly more metallic than the single field dwarf stars. This is a very important result suggesting that metallicity plays a very important role in the formation of massive planets independent of mass. When comparing the metallicity distributions of GP and SGP hosting giant stars  with the metallicity distribution of filed giant stars without planets, the KS statistics suggests probabilities of 0.007 (for GPH) and 0.105 (for SGPH) that the samples come from the same parent distribution. However, it is very important to note that while the GPH are more metallic than the field giant stars (the mean metallicity of the \citep{adibekyan_chemical_2012} + \citep{alves_determination_2015} sample is  $-$0.094 $\pm$ 0.007 dex), the giant hosts of SGP are slightly less metallic. This result suggest that giant planet (at least up to 4~M$_{\jupiter}$) formation around massive stars is more efficient if the disk is more metallic. Contrary, it seems that the most massive giant planets and/or BDs do not need metal-rich disks to be formed around massive stars.

Since our sample of planet hosting and single stars do not consist a single, well defined sample, it is not possible to determine the occurrence rate of very massive planets and test whether it depends on metallicity and stellar mass. Perhaps, a precise knowledge about a dependence (or absence of a dependence) of very massive planets and BD occurrence rate on stellar mass would help to better understand planet formation in massive disks. Different independent studies suggest that giant planet occurrence rate increases with the mass of the host star \citep[e.g.][]{reffert_precise_2015, jones_four_2016, ghezzi_retired_2018} for up to masses of about 2~M$_{\odot}$ and rapidly drops  for masses larger than about 2.5~M$_{\odot}$ \citep[e.g.][]{reffert_precise_2015}\footnote{For a discussion about the possibility of planet engulfment producing a lack of close-in planets orbiting intermediate mass stars see e.g. \citet{sato_planetary_2008, villaver_orbital_2009, kunitomo_planet_2011}.}. This dependence is well explained by planet formation models based on CA \citep[e.g.][]{ida_toward_2005, kennedy_planet_2008, mordasini_extrasolar_2012}. GI models also predict an increase of giant planet formation efficiency with stellar mass\footnote{GI predict formation of more massive and more protoplanets as the disk/stellar mass increases \citep[][]{boss_formation_2011, vorobyov_formation_2013}.} \citep[][]{boss_formation_2011, vorobyov_formation_2013, kratter_gravitational_2016}. However, the short lifetime of massive disks (for stars with intermediate masses)   affecting the formation efficiency of giant planets in the CA models should not be a problem for the giant planets formed through GI. Unfortunately, the limited number of BD detections does not permit to firmly  conclude whether BD occurrence rate is a function of stellar mass or not. The occurrence rate of relatively short-period (within a few years) BDs around solar-mass stars is estimated to be between  0.5\% and 1\% \citep[e.g.][]{grether_how_2006, sahlmann_search_2011, grieves_exploring_2017}. \citet{borgniet_extrasolar_2018} did not find any BD mass companion within about 1000 days around a sample of 225  AF-type main sequence stars observed with SOPHIE and/or HARPS spectrographs. These authors estimated the  upper limit of the occurrence rate of BDs around these stars to be below 4\%. An occurrence rate of 1.6\% for BDs around intermediate-mass giant and main-sequence stars was estimated by \citet[][]{jones_eccentric_2017}, which is slightly higher than the rate around lower mass stars. A similar occurrence rate of about 2\% for BDs orbiting white dwarfs (these are typically progeny of AF-type main sequence stars) was determined by \citet{girven_da_2011}\footnote{Note  that \citet{girven_da_2011} adopted an indirect detection method of BD based on the existence of near-infrared excess emission in the spectral energy distributions of white dwarfs.},  again giving a tentative evidence that BD formation might be more efficient around massive stars. Interestingly, the parent stars of the BD candidates detected in \citet[][]{jones_eccentric_2017} are metal-rich. According to the CA model of \citet{mordasini_extrasolar_2009} the formation of super-massive planets and BDs are possible in massive and metal-rich disks at large distances.

In summary, our data and results do not support the previous hints and claims \citep[e.g.][]{santos_observational_2017, narang_properties_2018, schlaufman_evidence_2018} about the existence of a breakpoint planetary mass at 4~M$_{\jupiter}$ above and bellow which planet formation channels are different. iant planet formation (independent of their mass) around solar-like stars preferentially occurs in metal-rich disks. These planets thus can be result of CA process or be formed by TD model of \citet{nayakshin_dawes_2017}. In contrast, it seems that the most massive planets and BDs orbiting massive stars can be formed regardless of the disk metallicity. Such planets are not typical outcome of CA models and it is natural to suggest that they might have been formed via GI which efficiently produce very massive planets in massive and metal-poor disks \citep{rafikov_can_2005, nayakshin_dawes_2017}. Interestingly,  several studies suggest that BDs with masses above and bellow $\sim$ 42~M$_{\jupiter}$ might have formed by different processes \citep[e.g.][]{sahlmann_possible_2011, ma_statistical_2014, mata_sanchez_chemical_2014, maldonado_searching_2017}. In particular, low-mass BDs can be formed by disk instability and the high-mass BDs via cloud fragmentation as stars \citep{ma_statistical_2014}.

\subsection{Low-mass planets and metallicity}

Ever since the first giant exoplanet was discovered orbiting a Sun-like star, the search has been ongoing for small, rocky planets around other stars, evocative of Earth and other terrestrial planets in the Solar System. Several ongoing and upcoming missions (e.g. TESS: \citet{ricker_transiting_2015}; CHEOPS \citet{fortier_cheops:_2014}; PLATO \citet{rauer_plato_2014})  and instruments (e.g. HARPS \citet{mayor_setting_2003}; HARPS-N \citet{cosentino_harps-n:_2012}; SOPHIE \citet{perruchot_sophie_2008}; CARMENES \citet{quirrenbach_carmenes_2014}; ESPRESSO \citet{pepe_espresso_2013}; SPIRou \citet{artigau_spirou:_2014}) are developed to search and characterize low-mass exoplanets, which should ultimately help us to understand the formation and evolution of these planets. However, the detection of low-mass/small-sized planets is not an easy task not only because of the very small photoelectric and RV signals, but also because of the activity signals coming from the host stars that often have the same magnitude as the planetary signal and can strongly perturb the detection of these planets and/or mimic a planetary signal \citep[e.g.][]{queloz_no_2001, oshagh_effect_2013, dumusque_earth-mass_2012, santos_harps_2014, hatzes_periodic_2016, faria_uncovering_2016, tal-or_carmenes_2018}. These difficulties not only limits the number of detected low-mass planets, but also make very hard to construct a control sample of stars without low-mass planets for RV surveys and make practically impossible\footnote{Transit method detects only planets which orbital planes are very close to the line of sight, thus all the stars with planets in inclined (with respect to the line of sight) orbital planes will appear as single stars.} for transit surveys \citep{buchhave_metallicities_2015}. Obviously, this makes  the comparison of the properties of stars with and without low-mass/small-sized planets very difficult. Such a comparison is crucial to constrain the models of low-mass planet formation and for understanding environmental conditions required for their formation. In fact no strong correlation between low-mass planet frequency and metallicity is predicted by most of the CA based models \citep[e.g.][]{mordasini_extrasolar_2012, hasegawa_planet_2014}. However, there are subtle differences in the relation between the frequency of these planets and metallicity predicted by different CA models. For example, recent N-body simulations of planet formation via pebble accretion by \citet{matsumura_n-body_2017} suggest that the formation efficiency of low-mass planets subtly depends on the stellar metallicitiy. Their simulations show that at high metallicities large number of low-mass planets can be formed, but majority of them leave the systems because of dynamical instabilities. In the CA model of \citet[][]{hasegawa_planet_2014} formation of low-mass planets is almost independent and their frequency slightly decrease towards a minimum (at [Fe/H] of about $-$0.2 dex) and then increases again with increasing metallicity. In their model low-mass planets are considered as failed cores of giant planets and the observed minimum in their frequency is related with the rise in the frequency of giant planets at that metallicity. Unlike most of the GI models \citep[e.g.][]{boss_giant_1997, galvagni_early_2014}, GI/TD model of \citet{nayakshin_tidal_2015} can also explain the formation of low-mass planets. The model of \citet{nayakshin_tidal_2015}, as the CA models, also predicts high occurrence rate of these planets at low metallicities. However, in their model the frequency of low-mass planets decreases at super-solar metallicities.

Despite the aforementioned difficulties many groups tried to study the metallicity distribution of low-mass stars. Based  only on a sample of seven short-period Neptune-mass ($M\sin i$ $<$ 21~M$_{\oplus}$) planets, \citet{udry_harps_2006} suggested that their formation efficiency may not depend on the host metallicity.  This hint was later supported by the results of \citet{sousa_spectroscopic_2008} who used a sample of 17 low-mass planets ($M\sin i$ $<$ 25~M$_{\oplus}$) orbiting FGK and M-type dwarf stars. With the increasing number of RV detected low-mass planets, several independent studies confirmed that these planets (with masses bellow 30-40~M$_{\oplus}$)  do not show any preference towards metal-rich FGK  \citep[e.g.][]{ghezzi_stellar_2010, mayor_harps_2009, sousa_spectroscopic_2011, mayor_harps_2011, jenkins_hot_2013, sousa_sweet-cat_2018} and M-type stars \citep[e.g.][]{rojas-ayala_metallicity_2012, neves_metallicity_2013, gaidos_they_2016, hobson_testing_2018}. However, it seems that the maximum mass of super-Earth/Neptune-class planets ($M\sin i$ $\lesssim$ 40~M$_{\oplus}$) shows a dependence on metallicity \citep{courcol_upper_2016, petigura_four_2017}. As for the massive planets, when studying the metallicity distributions of low-mass planet hosts, the chosen upper planetary mass limit varies from work to work, mostly depending on the sample size. The mass below which planet formation does not depend on metallicity (if there is such a limit) is still not well known. It is worth to note, that both observations \citep{mayor_harps_2011} and CA based models \citep[e.g.][]{mordasini_extrasolar_2012, brugger_metallicity_2018} suggest a strong minimum starting at about 30~M$_{\oplus}$ in the mass distribution of low-mass planets. 

\begin{figure}[H]
\centering
\begin{tabular}{c}
\includegraphics[width=7.5cm]{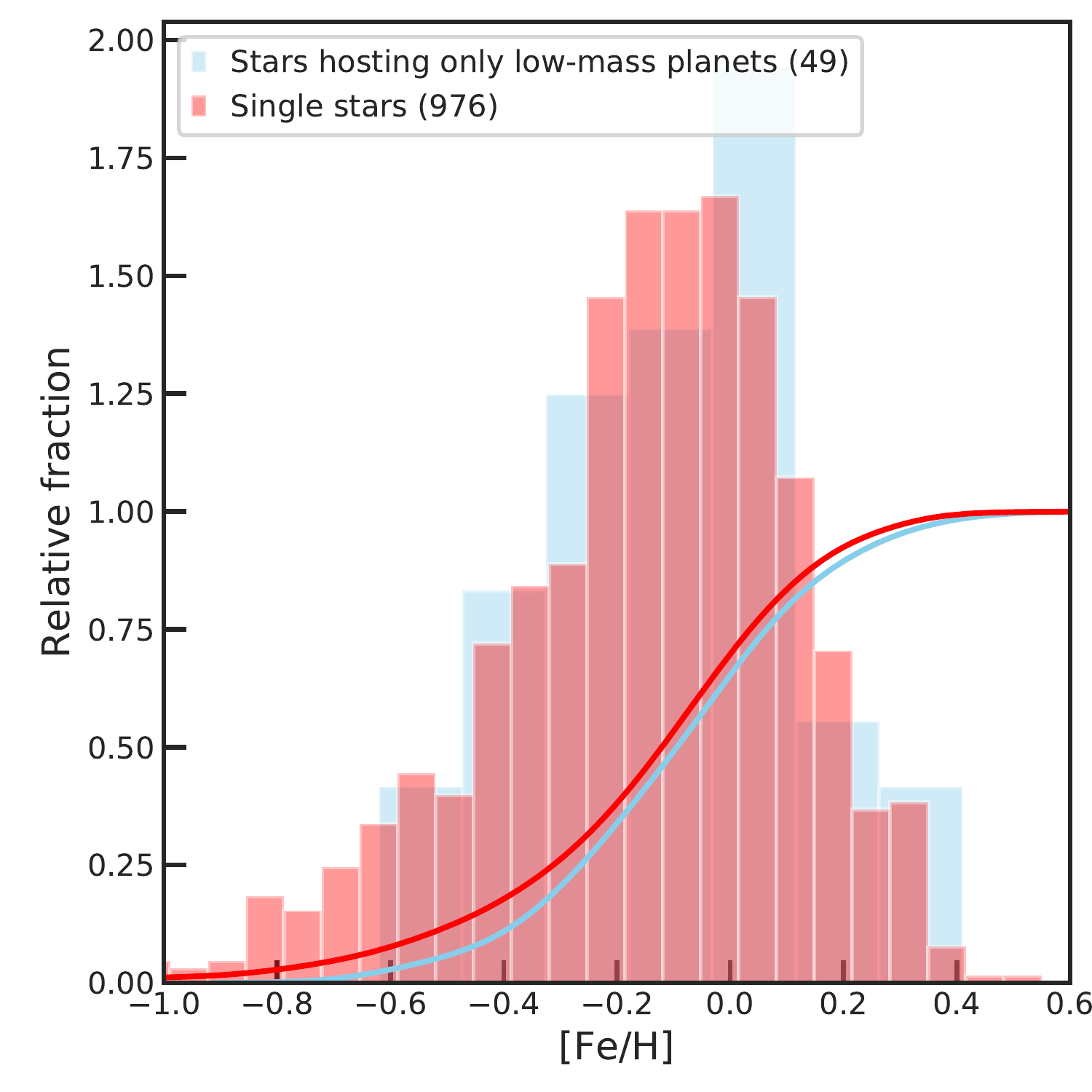}
\includegraphics[width=7.5cm]{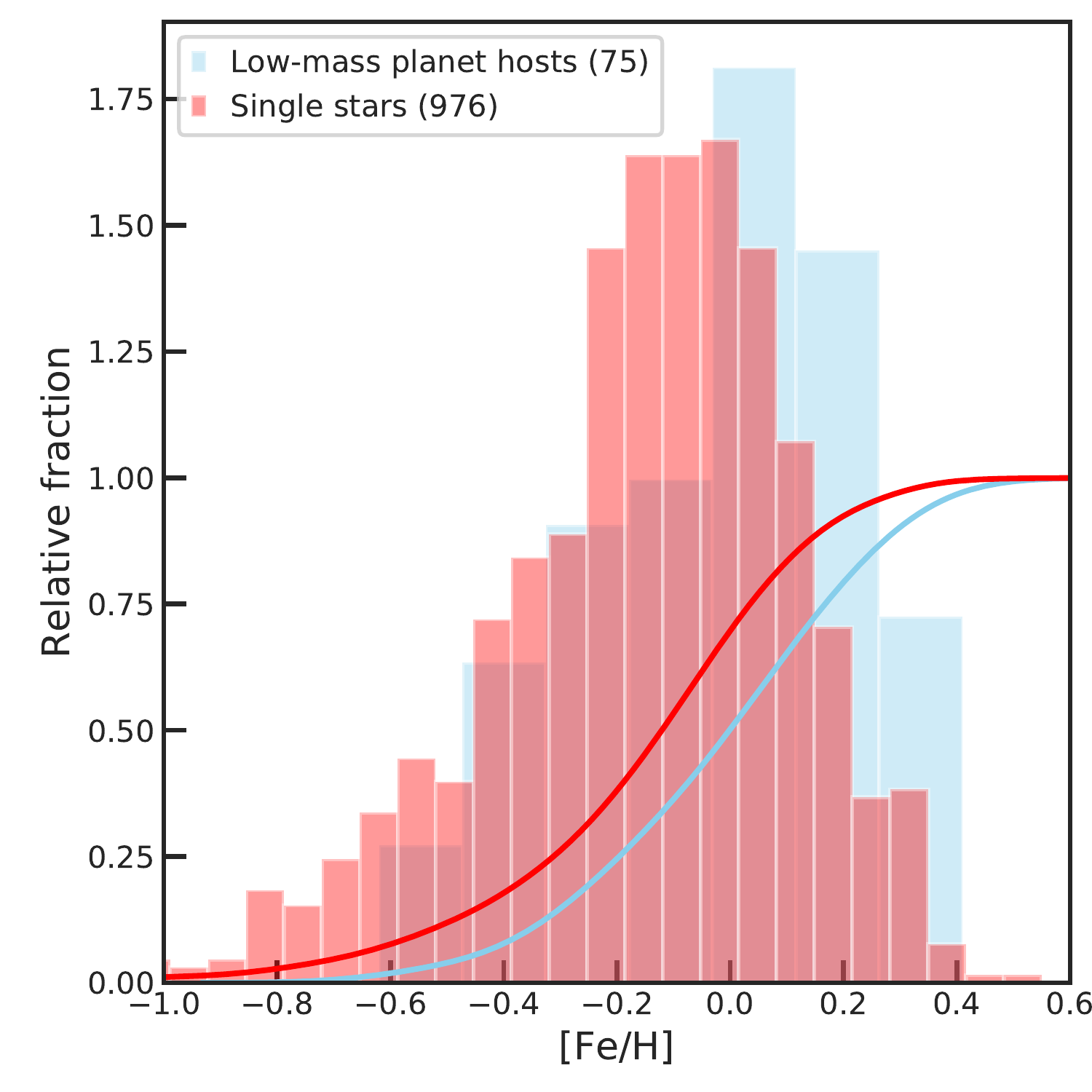}
\end{tabular}
\vspace{-0.2cm}
\caption{ Metallicity distributions of FGK type stars hosting exclusively low-mass planets (left panel, skyblue), stars hosting at least one low-mass planet (right panel, skyblue), and  stars without any detected planets (left and right panels, red). The KDE fit of the cumulative distribution of [Fe/H] for the sample of stars is also shown with the curves of corresponding colors.
}
\label{fig-low_mass_planet_metal}
\end{figure}

The left and right panels of Fig.~\ref{fig-low_mass_planet_metal} show the metallicity distributions of FGK type stars hosting planets or planetary systems with only low-mass detected planets ($<$ 30~M$_{\oplus}$) and systems where at least one of the planets have a mass of less than 30~M$_{\oplus}$. The comparison of these distributions with that of FGK type single dwarfs from the HARPS sample \citep{adibekyan_chemical_2012} shows  that the systems with only low-mass planets are less metallic ($-$0.101 $\pm$ 0.031 dex)  than the systems of low-mass planets accompanied by more massive planets ($-$0.021 $\pm$ 0.026 dex), but are slightly more metallic than the single stars ($-$0.159$\pm$0.009 dex). KS tests show that the difference in the metallicity distributions between stars with only low-mass planets and single stars is not statistically significant, while the metal excess of systems including low-mass planets when compared to the single stars is statistically significant (Table~\ref{table_KS_metal_HMPH}). As it is well known, and already discussed in this manuscript, giant planets tend to orbit stars with high metallicities. Therefore, exclusion of low-mass planet hosting systems that also contain giant planets will artificially reduce the number of metallic planetary systems with low-mass planets. However, the occurrence rate of giant planets is several times less than the occurrence rate of low-mass planets \citep[e.g.][]{mayor_harps_2011}. Thus, when using a well defined volume limited sample of stars with and without planets, where the fraction of low- and high-mass planet hosts reflect their occurrence rates, exclusion of stars hosting high-mass planets probably should not significantly affect the study of low-mass planet -- metallicity relation, at least for metallicities (e.g. [Fe/H] $\lesssim$ 0.1 dex) at which the giant planet frequency is not extremely high. When excluding super-metallic stars ([Fe/H] $>$ 0.1 dex) and comparing the metallicity distributions of single stars and low-mass planet hosts accompanied by massive planets, the KS tests provides a p-value of 0.17 that the two samples come from the same parent distribution. This test shows that the statistically significant difference in metallicity obtained for the single stars and stars hosting simultaneously low- and high-mass planets was due to the presence of super-metallic stars with giant planets. In Fig.~\ref{fig-low_mass_planet_metal} one can see that in our SWEET-Cat sample, the number of stars hosting only low-mass planets is 49 and the number of stars hosting low- and high-mass planets is 75-49 = 26. This relatively large number of multiple planetary systems with low-mass and giant planets is due to selection effects towards giant planets and the fact that about 90\% of cold giant planets accompanied by a low-mass planet \citep{zhu_super_2018}. 

Majority of the planets in the Earth-like to Neptune-like regime have been discovered via transit method (see Fig.~\ref{period_mass}). The large number of such planets detected by $Kepler$ provides a good base for a statistical studies of these small worlds. However, an important limitation of $Kepler$ planet studies is the faintness of their host stars which makes difficult to characterize them very accurately. Most of the early studies based on transiting planets reached to a conclusion that stars hosting exoplanets with radii smaller than about 4 R$_{\oplus}$ show wide range of metallicities, indistinguishable from the distribution of field stars without known planets \citep[e.g.][]{buchhave_abundance_2012, everett_spectroscopy_2013, buchhave_metallicities_2015}. Here I should remind again that a star without known planets does not necessarily mean that the star does not have planets.

Mass determination of the small-sized planets (R $<$ 4 R$_{\oplus}$) revealed the existence of two sub-populations: rocky and gaseous. A division between rocky planets and small-sized planets with gaseous envelopes was suggested to occur at about 1.6 R$_{\oplus}$ \citep{weiss_mass-radius_2014, rogers_most_2015}. Recently, \citet{fulton_california-kepler_2017} found a bimodal distribution for radii of Kepler short period ($<$ 100 d) small-sized planets with a clear lack of planets with radii between 1.5 and 2.0 R$_{\oplus}$. Later, it was shown that the position of the gap (sometimes called ``Fulton gap'') depends on orbital distance of the planets \citep{fulton_california-kepler_2018} and the mass/type of the hosts \citep{zeng_exoplanet_2017, fulton_california-kepler_2018}. In fact, \citet{jin_compositional_2018} showed that the location of the gap should depend on the composition of these planets. They argued that the observed gap is best explained by their cores being rocky (rather than icy) and that most close-in low-mass planets may be formed within the snow line. The presence of the gap is mostly explained by photoevaporaton of the atmospheres of these low-mass planets \cite[][but see also \citep{ginzburg_core-powered_2018,lehmer_rocky_2017} for alternative explanations]{owen_kepler_2013, lopez_understanding_2014, jin_planetary_2014, owen_evaporation_2017, fulton_california-kepler_2018}, which may depend on metallicity \citep{owen_metallicity-dependent_2018}. In addition, CA based \textit{in-situ} planet formation model of \citet{dawson_metallicity_2015} suggest that the presence or absence of gaseous atmosphere depends on the solid surface density of the protoplanetary discs where they have been formed. Similarly, \citet{owen_metallicity-dependent_2018} found that the  core masses  of low-mass planets are larger at high metallicities allowing to accreate larger H/He envelopes. If the photoevaporaton is the  mechanism producing the ``Fulton gap'' then the aforementioned results indicates that the metallicity distribution of the rocky planet hosts should be contaminated by the metallicities of the hosts of the photoevaporated cores that before the photoevaporaton had larger sizes.

Using a sample of about 400 $Kepler$ exoplanet candidate host stars with homogeneously derived metallicities \citet{buchhave_three_2014} found that there are three regimes of exoplanets - terrestrial-like planets (R $<$ 1.7 R$_{\oplus}$), gas dwarf planets (1.7 R$_{\oplus}$ $<$ R $<$ 3.9 R$_{\oplus}$), and ice or gas giant planets (R $>$ 3.9 R$_{\oplus}$) - that show different mean metallicities. In particular, \citet{buchhave_three_2014} showed that the terrestrial-like planets have a mean metallicity close to the solar value, while the hosts of gas dwarfs and ice/gas giants are more metallic than stars without known planets \citep{buchhave_metallicities_2015}. \citet{schlaufman_continuum_2015} using the sample of \citet{buchhave_three_2014} did not find a convincing evidence of the boundary at 1.7 R$_{\oplus}$, and argued that the data are better fitted by a continuous relation between radius and metallicity.  

Most of  the studies based on a relatively large number of transiting exoplanets with spectroscopically derived metallicities indicate that small-sized rocky planets (smaller than about 1.7 R$_{\oplus}$) show no preference towards high metallicities \citep[e.g.][]{buchhave_three_2014, mulders_super-solar_2016, narang_properties_2018, petigura_california-kepler_2018}, while the occurrence rate of larger transiting planets (1.7--3.9 R$_{\oplus}$) show a correlation with metallicity \citep[e.g.][]{buchhave_three_2014, mulders_super-solar_2016, petigura_california-kepler_2018}. By comparing the metallicity distribution of long-period ($>$ 25 days) small-sized (1--1.8~R$_{\oplus}$) planet hosts with the metallicity distribution of stellar hosts of planets with radii between 1 and 6 R~$_{\oplus}$ (independent of the orbital period) \citet{owen_metallicity-dependent_2018} found that the terrestrial-like planets are more common around low-metallicity stars and might have been formed  after the gas disc was dispersed. However, to confirm or infirm whether rocky planet formation is more prevalent around lower metallicity stars one needs to study the dependence of occurrence rate of this planets on metallicity. \citet[e.g.][]{petigura_california-kepler_2018} showed that the occurrence rate of warm super-Earths (P = 10--100 days and R = 1.0--1.7~R$_{\oplus}$) almost independent of metallicity, showing slight (insignificant) decrease with metallicity. However, it is important to note that  for transiting planets it is very hard to estimate the fraction of stars with planets \citep[][]{youdin_exoplanet_2011} and \citet[e.g.][]{petigura_california-kepler_2018}, as most of the works, used the average number of planets per star when studying the planet occurrence -- metalliciy dependence. Obviously, these two estimates of the occurrence rates are different (they would be the same if there was no star with more than one planet), and show different dependencies on stellar metallicity \citep{zhu_influence_2018}. Contrary to aforementioned results, \citet{wang_revealing_2015} suggested a universal planet--metallicity correlation for planet of all sizes. It is important to note that, while the stellar metallicites of their planet candidate hosts were derived spectroscopically, the metallicities of their control sample stars were based on photometric stellar parameters with further conversion to more ``representative'' stellar parameters \citep{wang_revealing_2015}. \citet{zhu_dependence_2016} suggested that the high occurrence rate and low detection efficiency of low-mass planets can explain the discrepant results regarding the small-sized planet -- metallicity correlations obtained by different authors.

When studying the metallicity dependence of low-mass/small size planets it is very important to take into account the relation between orbital periods of planets and their host stars metallicities \citep{adibekyan_orbital_2013, beauge_emerging_2013, dawson_metallicity_2015, adibekyan_which_2016, mulders_super-solar_2016, petigura_california-kepler_2018, wilson_elemental_2018, dong_lamost_2018}. In particular, \citet{adibekyan_orbital_2013} found that the super-Earth-like planets ($M\sin i$ $<$ 10~M$_{\oplus}$) orbiting metal-rich stars have orbital periods shorter than about 20 days, whereas planets orbiting metal-poor stars span a wide range of orbital periods \citep[see also][]{dawson_metallicity_2015, adibekyan_which_2016}. However, these results were contested by \citet{mulders_super-solar_2016} who argued that the observed trends might be due to selection effects. A systematic  excess of short period ($\lesssim$ 10 days) rocky planets ($<$ 1.7 R$_{\oplus}$) around metal-rich stars was reported in several works \citep{mulders_super-solar_2016, petigura_california-kepler_2018, wilson_elemental_2018}. The metallicity preference of hot rocky planets is explained by the possible dependence of efficient inward migration of solids on metallicity \citep{petigura_california-kepler_2018},  higher
rates of planet-planet scattering in metal-rich disks \citep{petigura_california-kepler_2018}, and/or dependence of planet trap at the inner age on metallicity \citep{mulders_super-solar_2016, wilson_elemental_2018}. An alternative possibility is that hot rocky planets and hot Jupiters might share the same formation mechanism \citep{mulders_super-solar_2016}.  

Despite aforementioned results suggesting that low-mass/small planet formation can be efficiently occur around stars with wide range of metallicities there seems to be an observational evidence that at a given metallicity these planets prefer to orbit stars with high [Mg/Si] abundance ratio \citep{adibekyan_stellar_2015, adibekyan_mg/si_2017} and at low-iron regime preferably orbit stars enhanced in $\alpha$-elements such as Mg and Si \citep{adibekyan_exploring_2012}. These results suggest that low-mass planet formation is not completely independent of the   composition of the protoplanetary disk.

\section{Summary and conclusion} \label{conclusion}

In this manuscript I reviewed and discussed the dependence of planet occurrence on stellar metallicity (iron abundance). In the upcoming manuscripts I will review i) the dependencies observed between exoplanet properties (e.g. orbital properties, multiplicity) and the metallicity (iron abundance) of their host stars (Paper II: 'Heavy metal rules II. Exoplanet properties and stellar metallicity') and ii) the role of individual heavy and light elements in the formation of exoplanets and the link between composition of exoplanets and their hosts stars (Paper III: 'Heavy metal rules III. Exoplanets and elements other than iron').

\subsection{Questions to answer or to think about}

Before starting to write this review, I was hoping that by gathering/collecting all the available information and evidences coming from observations and  theoretical predictions about formation of exoplanets of different types, I and/or the reader will get closer in answering to some important questions that I had in my mind. I hope the reader will be more successful in answering to this questions than the writer.

The origin of giant planet -- metallicity correlation. As discussed in the manuscript, formation of giant planets depends on several characteristics of the protoplanetary disks that are inter-correlated and all can produce a correlation between giant planet occurrence and metallicity. The most direct explanation of the correlation would go as, large amount of metals in the disk means large amount of material to build cores of giants which would translate into higher probability to form gas giants. However, at high metallicities the disk lifetime is longer \citep{ercolano_metallicity_2010} which makes easier (enough time) to build cores of giants and hence means higher probability to form gas giants as well. Besides the disk lifetime and metallicity, the mass of the disk also plays an important role for the formation of giant planets \citep[e.g.][]{kennedy_planet_2008}. As discussed in the manuscript, the disk mass, in turn correlates with the metallicity (due to selection effects) and disk lifetime \citep[e.g.][]{ribas_protoplanetary_2015}. The frequency of massive planets is lower than the frequency of systems with super-Earth-like planets. The latter systems, in principle, should have had enough material to build a core for giant planets. This may suggest that not the availability of core building blocks in the disk, but the conditions for planetary embryos to merge and acquire the critical core mass for giant planet formation is what might determine the formation probability of gas giants. \citet[][]{liu_migration_2016} suggested that the parameter determining these conditions  is the critical disk accretion rate, which correlates with metallicity and disk mass. Finally, as a very alternative explanation, it was proposed that the giant planet -- metallicity correlation can have a secondary origin \citet{haywood_correlation_2009}. Because of the radial metallicity gradient in the Galaxy \citep[e.g.][]{lemasle_galactic_2008,anders_red_2017}, the giant planet -- metallicity correlation can be easily recovered if the planet formation is more efficient in the inner Galaxy than in the solar neighborhood. \citet{haywood_correlation_2009} suggested that giant planet formation might be related to the surface density of molecular hydrogen H$_{2}$ in the Galaxy.

\textbf{Low-mass/small-sized planets and metallicity.} Until the recent advancement of the TD model \citep[e.g.][]{nayakshin_dawes_2017} terrestrial planet formation was explained exclusively by CA. Most of the  CA models and the TD model of \citet[][]{nayakshin_tidal_2015}\footnote{The typical outcome of most TD based models \citep[e.g.][]{forgan_towards_2013, galvagni_early_2014} are massive planets and BDs while terrestrial planets are formed very rarely and their survival is not guaranteed \citep{forgan_towards_2018}.} suggest practically no dependence of low-mass planet formation on disk metallicity, although still small differences exist in the predicted exact form of this dependence. The current observations seem to support the absence of correlation between the lowest-mass/smallest-sized planets and metallicity, at least for orbital periods longer than about 10 days \citep[e.g.][]{mulders_super-solar_2016, petigura_california-kepler_2018}. However, because of the very high frequency and low detectability of these planets it is still very difficult to firmly conclude whether terrestrial planet formation depends on metallicity or not \citep[e.g.][]{zhu_dependence_2016}. Very large samples of stars with and without planets will help to tackle this issue and help to select the model that explains the observations best.Hopefully such samples can be constructed in the near future with the ongoing and upcoming surveys and missions.

\subsection{The main new results and conclusions}

Besides making questions and extensive literature review about the role of metallicity on the formation of planets of different types, in this manuscript I also revisited most of the correlations related to metallicity of planet host stars reported in the literature. I used the SWEET-Cat database \citep{santos_sweet-cat:_2013} that provides homogeneously derived stellar parameters (including metallicities) for a very large sample of planet host stars, especially planets detected through RV method. Among the many findings discussed in the manuscript, I highlight two main new results and findings.

\begin{itemize}

 \item The study of the dependence of planet masses on the mean metallicity  of their hosts revealed (see Sect.~\ref{sub-Jupiters})  that the hosts of sub-Jupiter mass planets ($\sim$0.6 -- 0.9~M$_{\jupiter}$) are systematically less metallic than the hosts of the Jupiter-like planets.
The results seem to suggest that at high metallicities, the longer disk lifetime \citep{ercolano_metallicity_2010} and higher amount of planet building material \citep[e.g.][]{mordasini_extrasolar_2012}, allows a formation of more massive Jupiter-like planets than at lower metallicities. 

 \item Recently, several authors suggested that planets more massive than about 4~M$_{\jupiter}$ tend to orbit around low-metallicity stars and might have formed in a different way than Jupiter-mass planets with $<$ 4~M$_{\jupiter}$ \citep[e.g.][]{santos_observational_2017, schlaufman_evidence_2018}. The results of this study show that giant planets with masses above and bellow 4~M$_{\jupiter}$ orbiting solar-like stars are preferentially metal-rich, which does not support the previous hints about the different formation mechanisms of these two groups of planets. Formation of these planets and the observed metallicity trend can be explained by the CA models \citep[e.g.][]{mordasini_extrasolar_2012} more easily than by the GI and TD models \citep[e.g.][]{boss_giant_1997, nayakshin_dawes_2017}. The results also show statistically significant difference in the metallicity distributions of giant stars ($>$ 1.5~M$_{\odot}$) hosting planets with masses greater or less than 4~M$_{\jupiter}$. Perhaps, GI based models should be able to explain the formation of the most massive giant planets ($>$ 4~M$_{\jupiter}$) in low-meallicity environment more easily than CA based models. It is thus suggested that planets of the same mass can be formed through different channels depending on the disk mass i.e. environmental conditions.

\end{itemize}

Summarizing all the results and trends discussed in the manuscript one can conclude that there is no yet a single CA nor GI/TD model that can effectively explain the formation (and the observed trends with metallicity) of planets of all types. CA models can qualitatively explain the formation and observed trends with metallicity for all planets, but the super-massive planets ($>$ 4~M$_{\jupiter}$) orbiting metal-poor giant stars. GI can be responsible for the formation of these later planets, but will not explain the formation of low-mass planets and the giant planet -- metallicity correlation. The TD model of \citet[][]{nayakshin_dawes_2017} can qualitatively explain most of the metallicity trends discussed in this manuscript, but will fail in explaining the super-giant planet -- metallicity correlation observed for solar-type stars. 

The planet formation models constrained by the observations undergo significant improvements on how different physical processes are implemented. Observational exoplanetology does not stand still as well. Several ongoing and up-coming ground-based surveys and space-based missions (sometimes driven by theoretical predictions) are planned that will provide the exoplanet community with unprecedented amount of high quality data. Everything speaks about the bright future of exoplanetology. My very general conclusion, thus is that  \textit{heavy metal rules}!

%%%%%%%%%%%%%%%%%%%%%%%%%%%%%%%%%%%%%%%%%%
\vspace{6pt} 

%%%%%%%%%%%%%%%%%%%%%%%%%%%%%%%%%%%%%%%%%%
%% optional
%\supplementary{The following are available online at \linksupplementary{s1}, Figure S1: title, Table S1: title, Video S1: title.}

% Only for the journal Methods and Protocols:
% If you wish to submit a video article, please do so with any other supplementary material.
% \supplementary{The following are available at \linksupplementary, Figure S1: title, Table S1: title, Video S1: title. A supporting video article is available at doi: link.}

%%%%%%%%%%%%%%%%%%%%%%%%%%%%%%%%%%%%%%%%%%
%\authorcontributions{For research articles with several authors, a short paragraph specifying their individual contributions must be provided. The following statements should be used ``Conceptualization, X.X. and Y.Y.; Methodology, X.X.; Software, X.X.; Validation, X.X., Y.Y. and Z.Z.; Formal Analysis, X.X.; Investigation, X.X.; Resources, X.X.; Data Curation, X.X.; Writing—Original Draft Preparation, X.X.; Writing—Review \& Editing, X.X.; Visualization, X.X.; Supervision, X.X.; Project Administration, X.X.; Funding Acquisition, Y.Y.'', please turn to the \href{http://img.mdpi.org/data/contributor-role-instruction.pdf}{CRediT taxonomy} for the term explanation. Authorship must be limited to those who have contributed substantially to the work reported. }

%%%%%%%%%%%%%%%%%%%%%%%%%%%%%%%%%%%%%%%%%%
\funding{This work was supported by FCT - Fundação para a Ciência e a Tecnologia through national funds and by FEDER through COMPETE2020 - Programa Operacional Competitividade e Internacionalização by these grants: UID/FIS/04434/2013 \& POCI-01-0145-FEDER-007672; PTDC/FIS-AST/28953/2017 \& POCI-01-0145-FEDER-028953 and PTDC/FIS-AST/32113/2017 \& POCI-01-0145-FEDER-032113. I also acknowledge the support from FCT through Investigador FCT contract nr. IF/00650/2015/CP1273/CT0001.}

%%%%%%%%%%%%%%%%%%%%%%%%%%%%%%%%%%%%%%%%%%
\acknowledgments{I would like to thank the EXO-Earth team members for working together on the subject of this manuscript for many years. It is my special pleasure to thank Pedro Figueira, Jo\~{a}o Faria, Nuno Santos, Elisa Delgado Mean, S\'{e}rgio Sousa, Olivier Demangeon, Susana Barros, and Vardan Elbakyan for very interesting conversations and discussions. My double thanks to Mahmoud Oshagh for the very interesting  discussions and for inviting me to review this topic. I  also thank Sergei Nayakshin for the provided data and for very interesting discussion about the formation of planets through GI. I would also like to thank David Adibekyan for his critical comments that were sometimes difficult to figure out and to implement. Finally, I would like to thank  the referees for very constructive comments that helped to improve the quality of the manuscript.}

%%%%%%%%%%%%%%%%%%%%%%%%%%%%%%%%%%%%%%%%%%
\conflictsofinterest{The author declares no conflict of interest.}
%'' Authors must identify and declare any personal circumstances or interest that may be perceived as inappropriately influencing the representation or interpretation of reported research results. Any role of the funding sponsors in the design of the study; in the collection, analyses or interpretation of data; in the writing of the manuscript, or in the decision to publish the results must be declared in this section. If there is no role, please state ``The founding sponsors had no role in the design of the study; in the collection, analyses, or interpretation of data; in the writing of the manuscript, and in the decision to publish the results''.} 

%%%%%%%%%%%%%%%%%%%%%%%%%%%%%%%%%%%%%%%%%%
%% optional
\abbreviations{The following abbreviations are used in this manuscript:\\
\noindent 
\begin{tabular}{@{}ll}
RV  & radial velocity \\
CCD & Charge-Coupled Device \\
SS & Solar System \\
CA & core accretion\\
GI & gravitational instability \\
TD & Tidal Downsizing \\
ISM & interstellar medium \\
BD & brown dwarf \\
GTO & Guaranteed Time Observations \\
HARPS & High Accuracy Radial velocity Planet Searcher \\
HMPH & High Mass Planet Host \\
SnoP & Stars with not Planets \\
GPH & Giant Planet Host \\
GP & Giant Planet \\
SGPH & Super-Giant Planet Host \\
SGP & Super-Giant Planet \\
KS & Kolmogorov-Smirnov \\
std & standard deviation \\
TESS & Transiting Exoplanet Survey Satellite \\
PLATO & PLAnetary Transits and Oscillations of stars \\
AU & Astronomical Unit \\
KDE & Kernel Density Estimate
\end{tabular}}

%%%%%%%%%%%%%%%%%%%%%%%%%%%%%%%%%%%%%%%%%%
% Citations and References in Supplementary files are permitted provided that they also appear in the reference list here. 

% The following MDPI journals use author-date citation: Arts, Econometrics, Economies, Genealogy, Humanities, IJFS, JRFM, Laws, Religions, Risks, Social Sciences. For those journals, please follow the formatting guidelines on http://www.mdpi.com/authors/references
% To cite two works by the same author: \citeauthor{ref-journal-1a} (\citeyear{ref-journal-1a}, \citeyear{ref-journal-1b}). This produces: Whittaker (1967, 1975)
% To cite two works by the same author with specific pages: \citeauthor{ref-journal-3a} (\citeyear{ref-journal-3a}, p. 328; \citeyear{ref-journal-3b}, p.475). This produces: Wong (1999, p. 328; 2000, p. 475)

%=====================================
% References, variant B: external bibliography
%=====================================
\externalbibliography{yes}
\bibliography{references.bib}

%% for journal Sci
%\reviewreports{\\
%Reviewer 1 comments and authors’ response\\
%Reviewer 2 comments and authors’ response\\
%Reviewer 3 comments and authors’ response
%}

%%%%%%%%%%%%%%%%%%%%%%%%%%%%%%%%%%%%%%%%%%
\end{document}